\documentclass[aps,prstab,twocolumn,groupedaddress]{revtex4}

\usepackage{graphicx}
\usepackage{epstopdf}
\usepackage{dcolumn}
\usepackage{bm}
\usepackage{amsmath,amssymb}

\begin{document}

\preprint{}

\title{Accurate transfer maps for realistic beamline elements:\\Part I, straight elements}\thanks{Work supported by U.S. Department
of Energy Grant DE-FG02-96ER40949.}


\author{Chad E. Mitchell}
\email{cemitch@umd.edu}
\author{Alex J. Dragt}%
 \email{dragt@umd.edu}
\affiliation{%
Physics Department, University of Maryland, College Park, MD
}%

\date{\today}

\begin{abstract}
The behavior of orbits in charged-particle beam transport systems, including both linear and circular accelerators as well as final focus sections and spectrometers, can depend sensitively on nonlinear fringe-field and high-order-multipole effects in the various beam-line elements.  The inclusion of these effects requires a detailed and realistic model of the interior and fringe fields, including their high spatial derivatives.  A collection of surface fitting methods has been developed for extracting this information accurately from 3-dimensional field data on a grid, as provided by various 3-dimensional finite-element field codes.  Based on these realistic field models, Lie or other methods may be used to compute accurate design orbits and accurate transfer maps about these orbits.   Part I of this work presents a treatment of straight-axis magnetic elements, while Part II will treat bending dipoles with large sagitta.  An exactly-soluble but numerically challenging model field is used to provide a rigorous collection of performance benchmarks.
\end{abstract}


\maketitle

\section{\label{sec:intro}Introduction}

For the design of high-performance linear accelerators, synchrotrons, and storage and damping rings it is essential to have realistic electric and magnetic field information for the various beam-line elements in order to compute accurate design orbits and accurate high-order transfer maps about the design orbits.  There are similar requirements for other charged-particle beam transport systems such as final focus sections, high-resolution spectrometers, and high-resolution electron microscopes.

Realistic field data can be provided on a grid with the aid of various 3-dimensional finite element codes, sometimes spot checked against measured data.  But the computation of high-order transfer maps based on this data appears to pose an insurmountable problem: the calculation of high-order transfer maps requires a knowledge of high derivatives of the field data.  The direct calculation of high derivatives based only on grid data is intolerably sensitive to noise (due to truncation or round-off) in the grid data \cite{Hildebrand}.  We will see that this problem can be solved by the use of surface methods. The effect of numerical noise can be overcome by fitting field data on a bounding surface far from the beam axis and continuing inward using the Maxwell equations.  (We recall that interior fields are uniquely specified by their values on bounding 
surfaces.)  While the process of differentiation serves to amplify the effect of numerical noise, the process of continuing inward using the Maxwell equations is {\it smoothing}.  This smoothing is related to the fact that harmonic functions take their extrema on boundaries.  When using surface methods, all fits are made on such boundaries.  Therefore, if these fits are accurate, interior data based on these fits will be even more accurate.

In this paper we will devote our attention to magnetic beam-line elements.  Static electric beam-line elements, and static electric and magnetic beam-line elements such as velocity selectors (Wien filters), could be treated in a similar way.  For a treatment of RF cavities, see \cite{Abell}.

There are two  magnetic cases that it is convenient to handle separately: straight and curved.  For straight magnetic elements such as solenoids, quadrupoles, sextupoles, octupoles, etc., and wigglers, it is convenient to employ cylindrical surfaces.  For the case of curved magnetic elements, such as dipoles with large design-orbit sagitta, it is  convenient to employ the surface of a bent box with straight ends.  In all cases the bounding surface will surround the design orbit within the beam-line element and will extend into the fringe-field regions outside the beam-line element, thus taking into account all fringe-field effects as well as all effects within the body of the beam-line element.

For the case of straight beam-line elements it is convenient to describe the magnetic field in terms of a magnetic scalar potential $\psi$.   Then, if one wishes to compute transfer maps in terms of canonical coordinates, one can proceed with the aid of an associated vector potential $\mbox{\boldmath $A$}$ computed from $\psi$.  Alternatively, if one wishes to integrate noncanonical equations employing the magnetic field $\mbox{\boldmath $B$}$, it can be obtained from the relation $\mbox{\boldmath $B$} = \nabla \psi$.

For the case of curved beam-line elements it is convenient to work directly with the vector potential.  Its use in the case of canonical coordinates is then immediate.  If instead one wishes to integrate noncanonical equations employing the magnetic field $\mbox{\boldmath $B$}$, it can be obtained from the relation $\mbox{\boldmath $B$} = \nabla \times \mbox{\boldmath $A$}$.
 
In this paper we will treat the case of straight beam-line elements.    
For this purpose we will employ the surface of a (virtual) cylinder of uniform cross-section (circular, elliptical, or rectangular) surrounding the beam, and which lies within all iron or other magnetic sources.  In these cases, a Green function can be determined for the geometry of the fitted domain as a series composed of known, orthogonal, and complete  functions.  In the case of bent-bore magnetic elements, such as dipoles with large sagitta, it is not possible to obtain a suitable Green function.  New techniques have therefore been developed for treating cases of such general geometries, and these techniques will be presented in a subsequent paper.

Our emphasis will be on the accurate representation of fields in terms of scalar and vector potentials.  Once these representations are available, there are a variety of methods for computing the associated design orbits and transfer maps about these orbits.  An Appendix summarizes how this can be done when canonical coordinates  are employed and the associated Lie-algebraic structure is exploited.

A brief outline of this paper is as follows:  Section II reviews circular cylinder harmonic expansions and introduces a collection of functions, known as {\it on-axis gradients}, which uniquely characterize the magnetic scalar potential.  It also describes how the vector potential can be obtained from the scalar potential. The on-axis gradients themselves are generally unspecified functions of $z$.  In some simple cases they can be found analytically.  However, in general they must be determined numerically.  Section III describes how this can be done using known magnetic field values determined at points on some regular 3-dimensional grid.  In Section IV, we treat an analytically soluble model problem, that of a magnetic monopole doublet, which is used to benchmark these methods.  In Section V, we discuss the advantages of these surface-fitting methods that result from numerical smoothing.  In Section VI, we use these methods to study a proposed ILC damping ring wiggler.  The paper concludes with a summary and an Appendix.  Further detail may be found in \cite{Chad, Dragt2}.

\section{\label{sec:harmonic}Cylindrical Harmonic Expansions}

\subsection{\label{sec:harmonic:scalar} Scalar Potential}
In a current-free region the magnetic field $\mbox{\boldmath $B$}$ is curl free, and can therefore be
described most simply in terms of a magnetic scalar potential $\psi$.  Because $\mbox{\boldmath $B$}$ is also divergence free, $\psi$ must obey the Laplace equation 
\begin{equation}
\nabla^2 \psi=0. \label{laplace}
\end{equation}

Since we wish to describe straight beam-line elements, it is convenient to work initially in circular cylindrical coordinates $\rho$, $\phi$, and $z$ with 
\begin{equation}
x=\rho \cos \phi, \quad\quad y=\rho \sin \phi.  \label{xyeq}
\end{equation}
We note for future use that (\ref{xyeq}) can be written in the form
\begin{equation}
x+iy=\rho e^{i\phi}
\end{equation}
from which it follows that, for non-negative integers $l$ and $m$,
\begin{subequations}
\begin{equation}
\rho^{2\ell}=(x^2+y^2)^\ell,
\end{equation}
\begin{equation}
\rho^m\cos m\phi=\mathcal{R}e[(x+iy)^m],
\end{equation}
\begin{equation}
\rho^m\sin m\phi=\mathcal{I}m[(x+iy)^m].
\end{equation}
\end{subequations}
We see that {\em even}  powers of $\rho$ and the combinations $\rho^m\cos m\phi$ and $\rho^m\sin m\phi$ are {\em analytic} (in fact, polynomial) functions of $x$ and $y$.

A general solution $\psi$ satisfying the Laplace equation and analytic near the axis $\rho=0$ takes the form
\begin{align}
\psi(\rho,\phi,z)&=\sum_{m=0}^{\infty}\int_{-\infty}^{\infty}dk{\;} I_m(k\rho)e^{ikz}\left[G_{m,s}(k)\sin{m\phi}\right. \nonumber \\
&\left.+G_{m,c}(k)\cos{m\phi}\right].  \label{Laplacecylinder}
\end{align}
Note that the term containing $G_{0,s}(k)$ does not contribute to the above sum, and we may set $G_{0,s}=0$ without loss of generality.
By utilizing the Taylor series of the modified Bessel function $I_m$, we may write $\psi$ in the form of a circular cylinder harmonic (multipole) expansion:
\begin{equation}
\psi(\rho,\phi,z)=\sum_{m=0}^{\infty}\left[\psi_{m,s}(\rho,z)\sin{m\phi}+\psi_{m,c}(\rho,z)\cos{m\phi}\right], \label{cylharmonic}
\end{equation}
where for $\alpha=s$, $c$, 
\begin{equation}
\psi_{m,\alpha}(\rho,z)=\sum_{l=0}^{\infty}\frac{(-1)^lm!}{2^{2l}l!(l+m)!}C_{m,\alpha}^{[2l]}(z)\rho^{2l+m}.
\end{equation}
The functions $C_{m,\alpha}^{[n]}$, known as on-axis gradients \cite{Chad, Dragt2}, are defined by
\label{cylindergradients}
\begin{equation}
C_{m,\alpha}^{[n]}(z)=\frac{i^n}{2^m{m!}}\int_{-\infty}^{\infty}dk{\;}e^{ikz}k^{m+n}G_{m,\alpha}(k).\label{CfromG}
\end{equation}
Note that 
\begin{equation}
C_{m,\alpha}^{[n]}(z)=d^nC_{m,\alpha}^{[0]}(z)/d z^n.
\end{equation}

We will require an expansion of (\ref{cylharmonic}) in the transverse variables $x$ and $y$.
Define the polynomials
\begin{subequations}
\begin{align}
F_s^{l,m}=(x^2+y^2)^l\mathcal{I}m(x+iy)^m, \\
F_c^{l,m}=(x^2+y^2)^l\mathcal{R}e(x+iy)^m,
\end{align}
\end{subequations}
for integer $l\geq 0$ and $m\geq 0$.  In Cartesian coordinates we then have:
\begin{align}
\psi(x,y,z)&=\sum_{l=0}^{\infty}\sum_{m=0}^{\infty}\frac{(-1)^lm!}{2^{2l}l!(l+m)!}\left[C_{m,s}^{[2l]}(z)F_s^{l,m}(x,y)\right. \nonumber \\
&\left. +C_{m,c}^{[2l]}(z)F_c^{l,m}(x,y)\right].  \label{psieq}
\end{align}
Note that each polynomial $F_{\alpha}^{l,m}$ is homogeneous of degree $2l+m$. 

\subsection{\label{sec:harmonic:vector} Vector Potential}
To determine an associated vector potential $\mbox{\boldmath $A$}$, we must find a solution to the coupled system of equations $\nabla\times\mbox{\boldmath $A$}=\nabla\psi$, where $\psi$ is given by the series (\ref{psieq}).  We also must select some particular gauge.  And even after a particular type of gauge has been chosen, say a Coulomb gauge, there is still some remaining freedom \cite{Dragt2}.  One convenient Coulomb gauge choice is given by the rules
\begin{widetext}
\begin{subequations}\label{Atran}
\begin{align}
A_x&=\sum_{l=0}^{\infty}\sum_{m=0}^{\infty}(-1)^l\frac{m!}{2^{2l+1}l!(l+m+1)!}\left[C_{m,s}^{[2l+1]}(z)F_c^{l,m+1}-C_{m,c}^{[2l+1]}(z)F_s^{l,m+1}\right], \\
A_y&=\sum_{l=0}^{\infty}\sum_{m=0}^{\infty}(-1)^l\frac{m!}{2^{2l+1}l!(l+m+1)!}\left[C_{m,c}^{[2l+1]}(z)F_c^{l,m+1}+C_{m,s}^{[2l+1]}(z)F_s^{l,m+1}\right], \\
A_z&=\sum_{l=0}^{\infty}\sum_{m=0}^{\infty}(-1)^l\frac{m!}{2^{2l}l!(l+m)!}\left[C_{m,c}^{[2l]}(z)F_s^{l,m}-C_{m,s}^{[2l]}(z)F_c^{l,m}\right].
\label{Along}
\end{align}
\end{subequations}
\end{widetext}
The coefficients $C^{[n]}_{m,s}(z)$ and $C^{[n]}_{m,c}(z)$ describe normal and skew components, respectively.  For example, in the {\em body} of a long normal dipole $(m=1)$, we expect $C^{[0]}_{1,s}(z)$ will be nearly constant (independent of $z$) and therefore $C^{[n]}_{1,s}(z)\simeq0$ for $n>0$.  Correspondingly, in the body of a long normal dipole, use of (\ref{Atran}) gives the results $A_x\simeq A_y\simeq0$ and
\begin{equation}
A_z=-C^{[0]}_{1,s}x.
\end{equation}
Similarly, in the body of a long normal quadrupole $(m=2)$, use of (\ref{Atran}) gives the results $A_x\simeq A_y\simeq0$ and
\begin{equation}
A_z=-C^{[0]}_{2,s}(x^2-y^2).
\end{equation}

For any set of analytic functions $C^{[n]}_{m,\alpha}(z)$ employed in (\ref{psieq}), the vector potential defined by (\ref{Atran}) satisfies the conditions:  1) $\nabla\times\mbox{\boldmath $A$}=\nabla\psi=\mbox{\boldmath $B$}$, 2) $\nabla\times\nabla\times\mbox{\boldmath $A$}=0$, and 3) $\nabla\cdot\mbox{\boldmath $A$}=0$.  Note that both Maxwell's equations and the Coulomb gauge condition are satisfied by construction.  In the following section, we will see how the coefficient functions $C_{m,\alpha}^{[n]}(z)$ can be numerically determined.

\section{\label{sec:surface} Surface Methods}
There are cases in which Taylor expansions of the form (\ref{psieq}) and (\ref{Atran}) can be found analytically.  
In general, however, we have available only measured or numerical three-dimensional magnetic field data on a discrete mesh of points distributed throughout the region of interest.  The required high derivatives of $\psi$ or $\mbox{\boldmath $A$}$ cannot be reliably computed directly from this data by numerical differentiation due to numerical noise, whose effect becomes worse with the order of derivative desired.  The effect of numerical noise, and its amplification by numerical differentiation, can be overcome by fitting on a bounding surface far from the axis and interpolating inward using the Maxwell equations.  Surface fitting methods have several advantages, including:

\begin{enumerate}

\item Only functions with known (othonormal) completeness properties and known (optimal) convergence properties are employed.

\item The Maxwell equations are exactly satisfied.

\item The results are manifestly analytic in all variables.

\item The error is globally controlled. Fields that are {\em harmonic} (fields that satisfy the Laplace equation) take their extrema on boundaries.  Both the exact and computed fields are harmonic.  Therefore their difference, the error field, is also harmonic, and must take its extrema on the boundary. But this is precisely where a controlled fit is made. Thus, the error on the boundary is controlled, and the interior error must be even smaller.

\item Because harmonic fields take their extrema on boundaries, interior values inferred from surface data are relatively insensitive to errors/noise in the surface data.  Put another way, the inverse Laplacian (Laplace Green function), which relates interior data to surface data, is {\em smoothing}. It is this smoothing that we seek to exploit.  In general, the sensitivity to noise in the data decreases rapidly (as some high inverse power of distance) with increasing distance from the surface, and this property improves the accuracy of the high-order interior derivatives needed to compute high-order transfer maps. 
\end{enumerate}

Let us briefly compare this approach to that of on-axis or midplane fitting.  In the case of on-axis fitting, it is common to use various analytic model profiles, such as {\em Enge} functions, and then differentiate them repeatedly to achieve objective 2 by continuing outward using the Maxwell equations.  However, these functions do not have completeness properties, item 1.  And there is no smoothing, item 5, to overcome the amplification of noise due to numerical differentiation. 

In the case of midplane fitting, one approach would be to attempt to employ expressions that relate the on-axis gradients to midplane data and its derivatives.  For example, in the case of midplane symmetry, there are the relations
\begin{subequations}
\begin{align}
&C^{[0]}_{m,c}(z)=0, \\
&C^{[0]}_{1,s}(z)=B_y(x=0,y=0,z), \\
&C_{2,s}^{[0]}(z)=\frac{1}{2}\frac{\partial B_y}{\partial x}\bigg|_{(0,0,z)}, \\
&C_{3,s}^{[0]}(z)=\frac{1}{6}\frac{\partial^2B_y}{\partial x^2}\bigg|_{(0,0,z)}
+\frac{1}{24}\frac{\partial^2B_y}{\partial z^2}\bigg|_{(0,0,z)}, \rm{etc.}
\end{align}
\end{subequations}
By repeatedly differentiating these relations with respect to $z$, one can obtain the $C^{[n]}_{m,s}(z)$ for $n>0$.  In general, the determination of $C^{[n]}_{m,s}(z)$ requires the computation of $m+n-1$ derivatives.  Although this approach achieves objective 2, since all relevant quantities are subsequently computed in terms of on-axis gradients, in the case where data is available only at grid points it presupposes the ability to compute very high-order derivatives by high-order numerical differentiation.  This is generally impossible, due to the high noise sensitivity associated with high-order numerical differentiation, because there is no intrinsic smoothing, item 5.

Another approach is to use an analytic functional form, with free parameters, that is known to satisfy the the 3-dimensional Laplace equation for all parameter values.  These parameters can be adjusted so that the field derived from this representation well approximates the field at various grid points.  (These grid points could be in the midplane, but could be out of the midplane as well.)  This representation can then be repeatedly differentiated to provide the required field derivatives.  However, commonly this fitting procedure has no known completeness/convergence property, item 1.  In some cases Fourier series representations with known completeness properties are used.  But with Fourier series representations, an artificial periodicity is imposed in the transverse horizontal direction.  As a result, the Fourier coefficients for the field expansions, call them $a_n$, can fall off at best as $(1/n^2)$[4].  Correspondingly, the Fourier coeficients in the associated expansion for $\psi$ fall off at best as $(1/n^3)$.  As a result, repeated differentiation produces nonconvergent series, and there is no analyticity, item 3.  Whatever representation is used, there is again no intrinsic smoothing to overcome the amplification of noise due to numerical differentiation.

\subsection{\label{sec:surface:circ} Use of Field Data on Surface of Circular Cylinder}

All three-dimensional electromagnetic codes calculate all three components of the field on some three-dimensional grid.  Also, such data is in  principle available from actual field measurements.  In this subsection we will describe how to compute the on-axis gradients from  such field data using the surface of a circular cylinder \cite{Venturini}.  Once these gradients are known, we may use (\ref{psieq}) and (\ref{Atran}) to compute the associated scalar and vector potentials.

\begin{figure}[hpt]
  \centering
 \resizebox{8.6cm}{!}{\includegraphics*{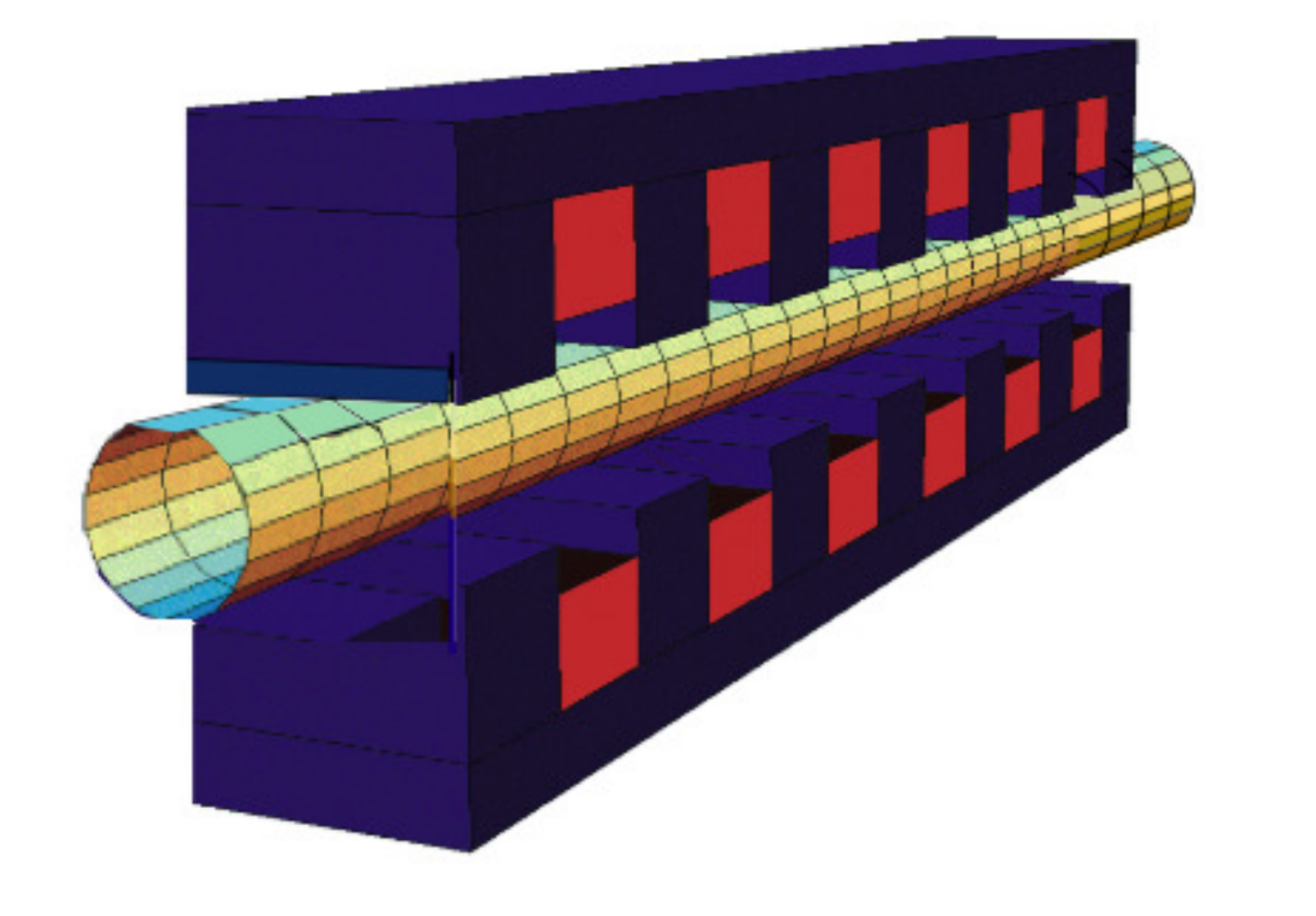}}
  \caption{\label{cylindergraphic} A circular cylinder of radius $R$, centered on the $z$-axis, fitting within the bore of a beam-line element, in this case a wiggler,  and extending beyond the fringe-field regions at the ends of the beam-line element.}
\end{figure}

Consider a circular cylinder of radius $R$, centered on the $z$-axis, fitting within the bore of the beam-line element in question, and extending beyond the fringe-field regions at the ends of the beam-line element. The beam-line element could be any straight element such as a solenoid, quadrupole, sextupole, octupole, etc., or it could be wiggler with little or no net bending. See Fig. \ref{cylindergraphic}, which illustrates the case of a wiggler.  

Suppose the magnetic field $\mbox{\boldmath $B$}(x,y,z)=\mbox{\boldmath $B$}(\rho,\phi,z)$ is given on a grid, and this data is interpolated onto the surface of the cylinder using values at the grid points near the surface.  Next, from the values on the surface, compute $B_\rho(R,\phi,z)$, the component of $\mbox{\boldmath $B$}(\rho,\phi,z)$ {\em normal} to the surface.  The major remaining task is to compute the on-axis gradients from a knowledge of $B_\rho(R,\phi,z)$.  See Fig. 2.  At this point we note that the functions $\exp(ikz)\sin(m\phi)$  and $\exp(ikz)\cos(m\phi)$ form a complete set over the surface of the circular cylinder.

\begin{figure*}[htp]
  \centering
\includegraphics*[height=7.5in,angle=0]{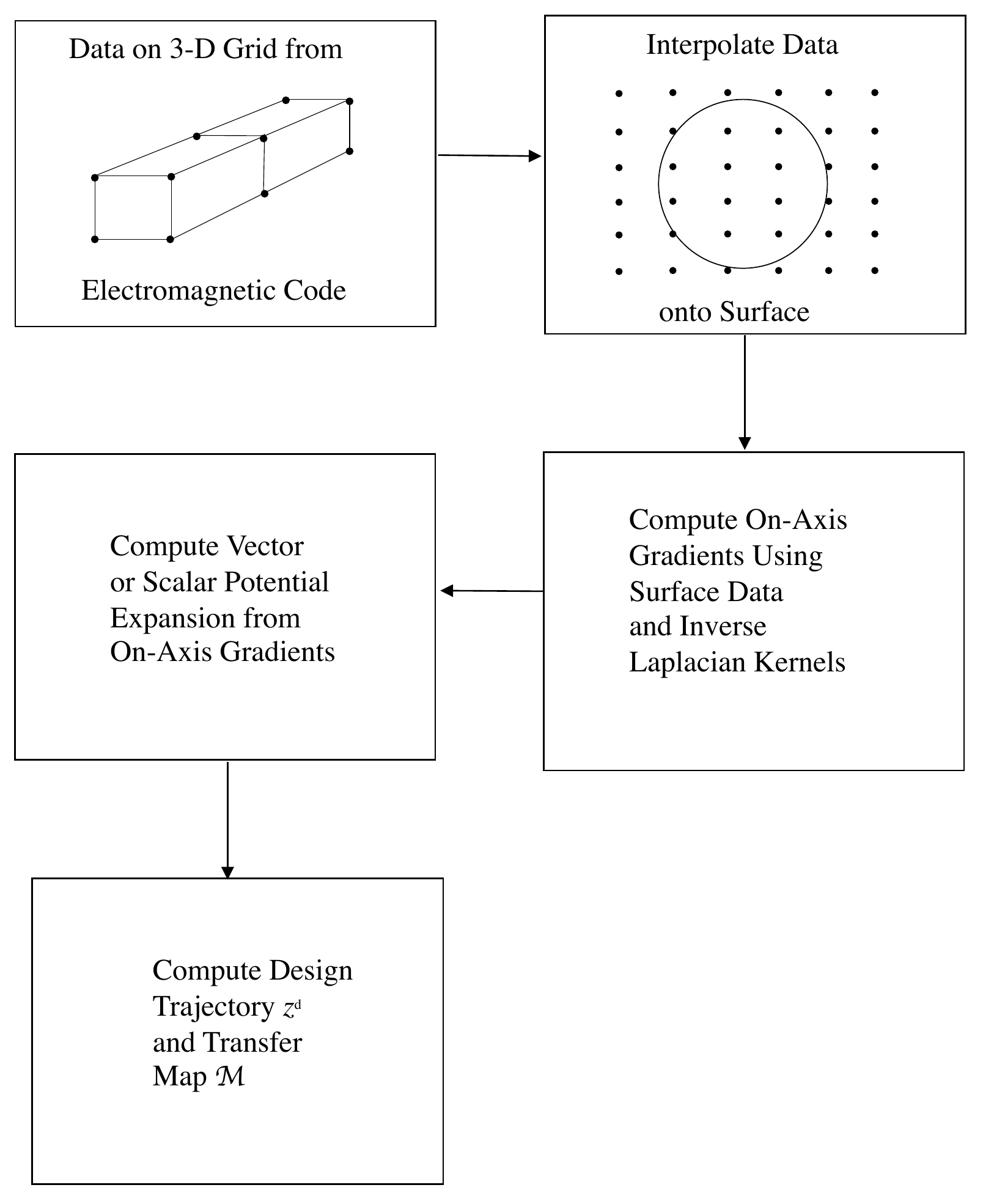}
  \caption{\label{algorithm} Calculation of realistic design orbit $z^d$ and its associated realistic transfer map $\cal M$ based on data provided on a 3-dimensional grid for a real beam-line element. Only a few points on the 3-dimensional grid are shown. In this illustration, data from the 3-dimensional grid is interpolated onto the surface of a cylinder with circular cross section, and this surface data is then processed to compute the design orbit (trajectory) and the associated transfer map about the design orbit.  The use of other surfaces is also possible, and may offer various advantages.}
\end{figure*}

Let $\tilde{B}_{\rho}(R,\phi,k)$ be the Fourier transform of $B_{\rho}(R,\phi,z)$ given by the integral
\begin{equation}
\tilde{B}_{\rho}(R,\phi,k)=\frac{1}{2\pi}\int_{-\infty}^{\infty}dz{\;} e^{-ikz}B_{\rho}(R,\phi,z).
\label{transform}
\end{equation}
Because $\mbox{\boldmath $B$}$ decays rapidly in the fringe field region, $B_{\rho}(R,\phi,z)$ is absolutely integrable along the $z$-axis, and therefore its Fourier transform is well defined.
Next define the functions $\tilde{b}_{m,s}$ and $\tilde{b}_{m,c}$ by
\begin{subequations}\label{intermcyl}
\begin{align}
\tilde{b}_{m,s}(R,k)&=\frac{1}{2\pi^2}
\int_{0}^{2\pi} d\phi \sin{m\pi}\int_{-\infty}^\infty dz{\;} e^{-ik z}
 B_\rho(R,\phi,z), \\
\tilde{b}_{m,c}(R,k)&=\frac{1}{2\pi^2}
\int_{0}^{2\pi} d\phi \cos{m\pi} \int_{-\infty}^\infty dz{\;} e^{-ik z}
B_\rho(R,\phi,z),
\end{align}
for $m\geq 1$ and
\begin{equation}
\tilde{b}_{0,c}(R,k)=\frac{1}{4\pi^2}
\int_{0}^{2\pi} d\phi \int_{-\infty}^{\infty} dz{\;} e^{-ikz}B_{\rho}(R,\phi,z).
\end{equation}
\end{subequations}
We know that
\begin{equation}
B_\rho(R,\phi,z)=
\left.\frac{\partial \psi(\rho,\phi,z)}{\partial \rho}\right|_{\rho=R},
\end{equation}
from which it follows, using the representation (\ref{Laplacecylinder}), that
\begin{align}
B_\rho(R,\phi,z)&=\sum_{m=0}^{\infty}\int_{-\infty}^{\infty}dk{\;} kI^{\prime}_m(k\rho)e^{ikz}\left[G_{m,s}(k)\sin{m\phi}\right. \nonumber \\
&\left.+G_{m,c}(k)\cos{m\phi}\right].  \label{cylfield}
\end{align}
Now substitute (\ref{cylfield}) into the right sides of (\ref{intermcyl}) and perform the indicated integrations to get the results
\begin{equation}
{\tilde{b}_{m,\alpha}}(R,k)=
G_{m,\alpha}(k)k I_{m}^\prime(k R),
\end{equation}
from which it follows that
\begin{equation}
G_{m,\alpha}(k)=
\frac{{\tilde{b}}_{m,\alpha}(R,k)}{k I^\prime_m(kR)}.
\end{equation}
This relation for $G_{m,\alpha}(k)$ can be employed in (\ref{CfromG}) to give the result
\begin{equation}
C_{m,\alpha}^{[n]}(z)=
\frac{i^n}{2^m m!}\int_{-\infty}^\infty dk{\;} e^{ikz}
k^{n+m-1}\frac{{\tilde{b}}_{m,\alpha}(R,k)}{I^\prime_m(kR)}.  \label{circulargradients}
\end{equation}

We have found an expression for the on-axis gradients in terms of field data (normal component) on the surface of the cylinder.  
Equation (\ref{circulargradients}) may be viewed as the convolution of Fourier surface data ${\tilde{b}}_{m,\alpha}(R,k)$ with the {\em inverse Laplacian} kernel $k^{n+m-1}/I^{\prime}_{m}(kR)$.  Moreover, this kernel has a very desirable property.  The Bessel functions 
$I^{\prime}_m(kR)$ have the asymptotic behavior
\begin{equation}
|I^\prime_m(kR)|\sim\exp(|k|R)/\sqrt{2\pi|k|R}{\;}{\rm as}{\;}
|k|\rightarrow \infty.
\end{equation}
Since $I^{\prime}_m(kR)$ appears in the denominator of (\ref{circulargradients}), we see that the integrand is exponentially damped for large $|k|$.  Now suppose there is uncorrelated point-to-point noise in the surface data.  Such noise will result in anomalously large $|k|$ contributions to the $\tilde{b}_{m,\alpha}(R,k)$.  But, because of the exponential damping arising from $I^{\prime}_m(kR)$ in the denominator, the effect of this noise is effectively filtered out. Moreover, this filtering action is improved by making $R$ as large as possible. This filtering, or {\em smoothing}, feature will be discussed in more detail in Section V.

We close this subsection with the remark that if one wishes to extract the $C^{[n]}_{0,c}(z)$ (monopole) on-axis gradients from field data, as is required for example in the case of a solenoid, it may be preferable to use the longitudinal component $B_z(R,\phi,z)$ on the surface of the cylinder rather than the normal component $B_\rho(R,\phi,z)$ \cite{Dragt2}.

\subsection{\label{sec:surface:ell}Use of Field Data on Surface of Elliptical Cylinder}

\subsubsection{\label{sec:surface:ell:bkgd}Background}

\begin{figure}[hpt]
  \centering
 \resizebox{8.6cm}{!}{  \includegraphics*{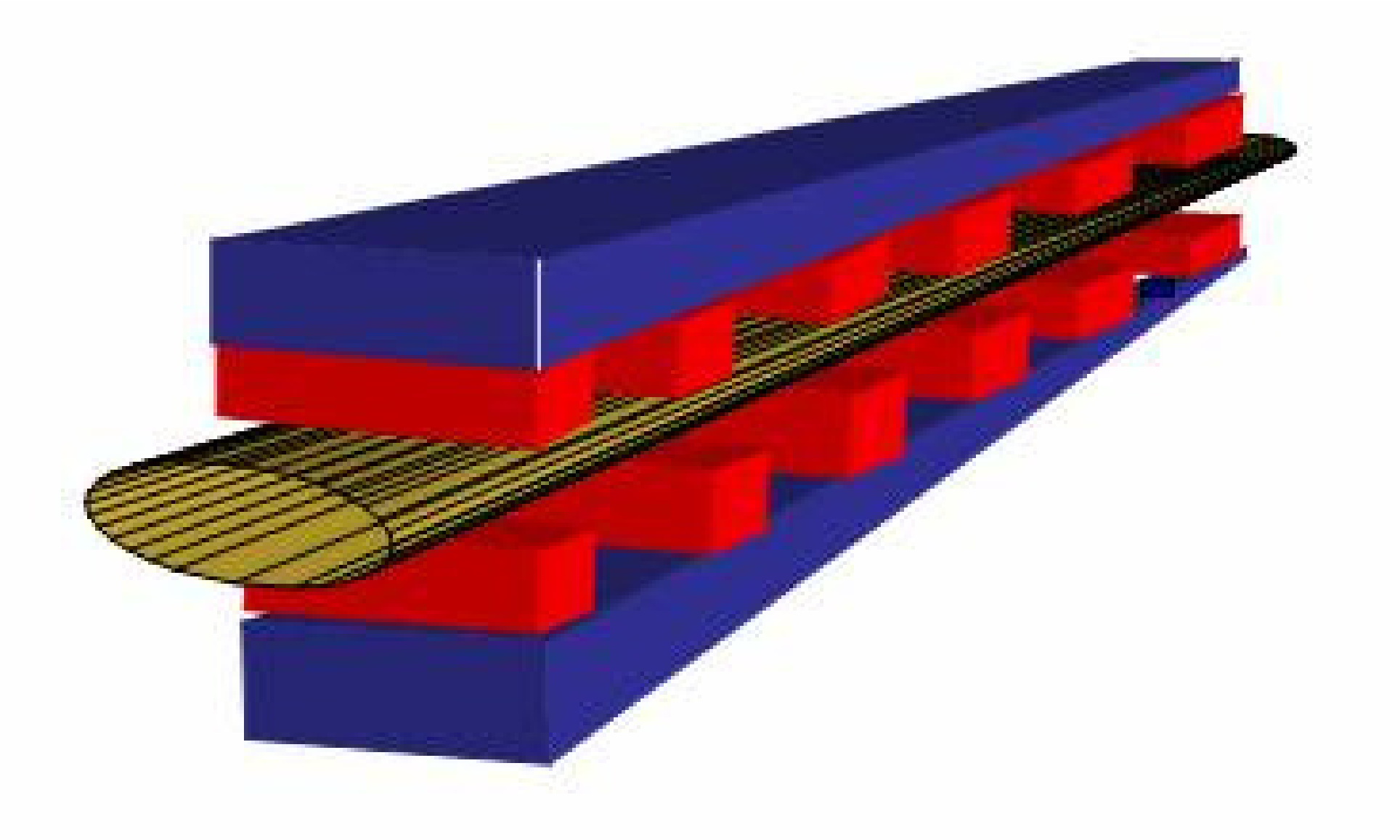}}
  \caption{\label{ellipticalcylindergraphic} An elliptical cylinder, centered on the $z$-axis, fitting within the bore of a wiggler, and extending beyond the fringe-field regions at the ends of the wiggler.}
\end{figure}

In the previous subsection we employed a cylinder with circular cross section, and observed mathematically that it is desirable for error insensitivity to use a cylinder with a large radius $R$.   Physically, this is because we want the field data points employed to be as far from the axis as possible since the effect of inhomogeneities (noise) in the data decays with distance from the inhomogeneity. Evidently the use of a large circular cylinder is optimal for beam-line elements with a circular bore.  However, for dipoles or wigglers with small gaps and wide pole faces, use of a cylinder with elliptical cross section should give improved error insensitivity.  See Fig. \ref{ellipticalcylindergraphic}.  In this subsection we will set up the machinery required for the use of elliptical cylinders, and apply it to the calculation of on-axis gradients based on field  data \cite{Chad, Dragt2}. 

\subsubsection{\label{sec:surface:ell:coord}Elliptic Coordinates}

\begin{figure}[hpt]
  \centering
   \resizebox{8.7cm}{!}{ \includegraphics*{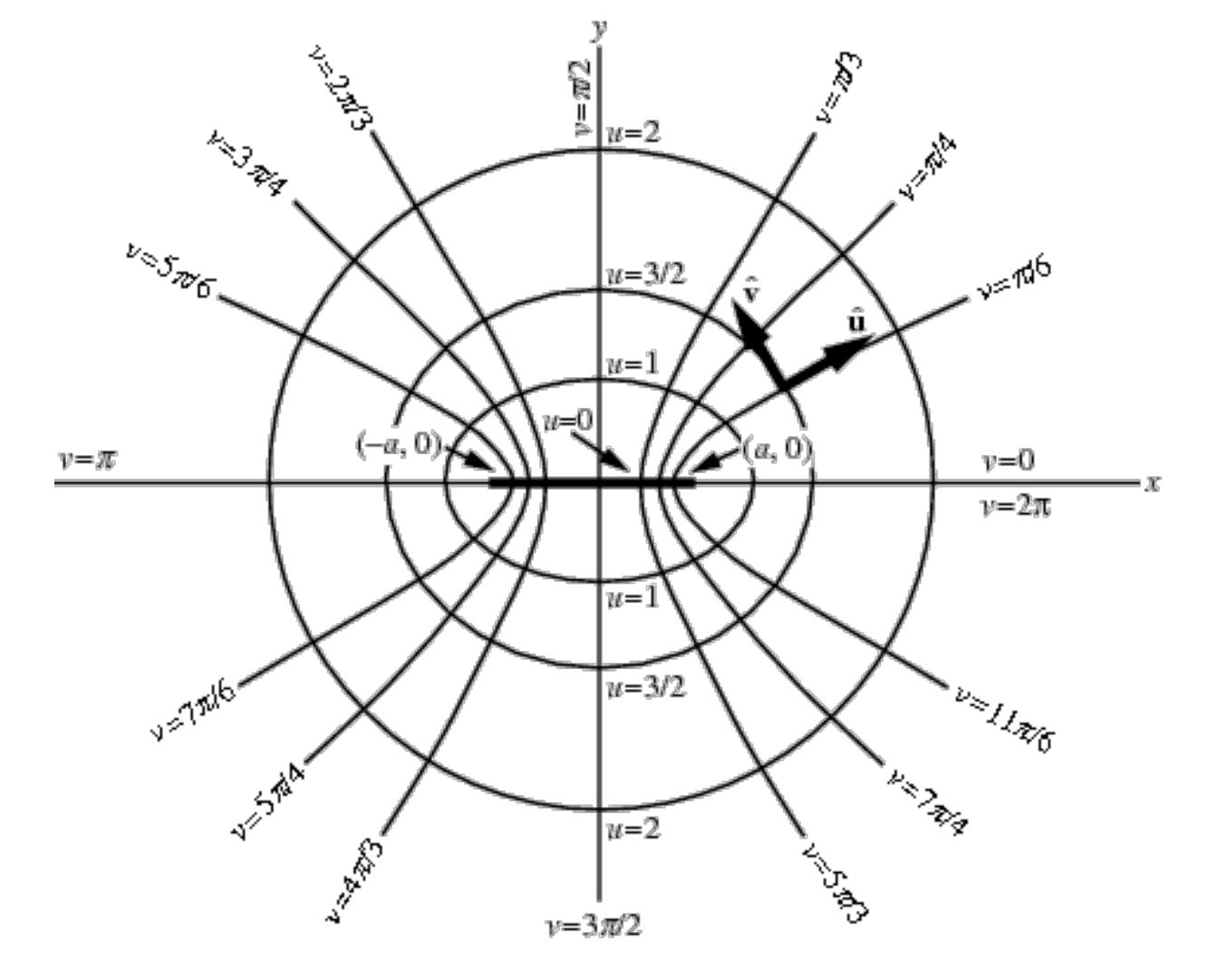}}
  \caption{\label{ellcoords} Elliptical coordinates showing contours of constant $u$ and constant $v$ \cite{Ellfigure}.  The foci are at $(\pm a,0)$, and in our case $a=f$.}
\end{figure}

\begin{figure}[hpt]
  \centering
 \resizebox{7.6cm}{!}{  \includegraphics*{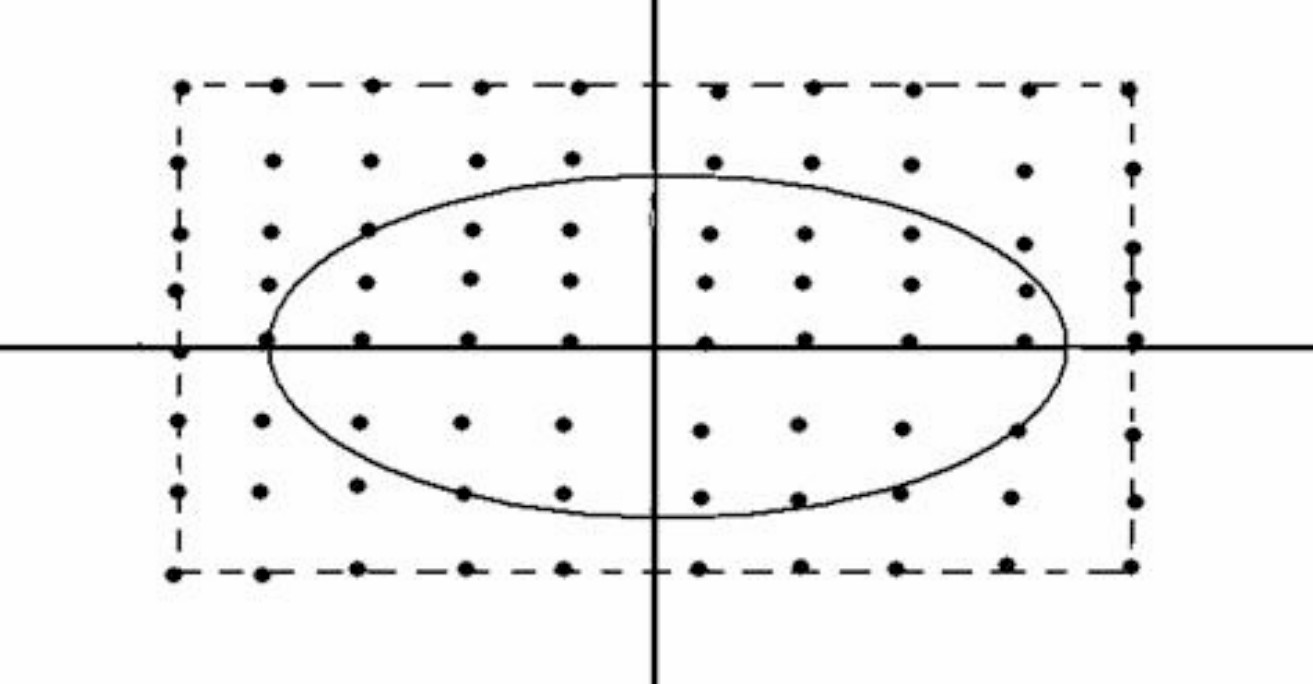}}
  \caption{\label{mesh} A square or rectangular grid in the $x$,$y$ plane for a fixed $z$ value on the 3-dimensional grid.  Values at data points near the ellipse are to be interpolated onto the ellipse.}
\end{figure}

Elliptic coordinates in the $x,y$ plane are described by the relations \cite{Feshbach}
\begin{subequations}\label{ellcoord}
\begin{align}
x&=f\cosh(u) \cos(v),\\
y&=f\sinh(u) \sin(v).
\end{align}
\end{subequations}
Contours of constant $u$, with $u\in[0,\infty]$, are nested ellipses with common foci located at $(x;y)=(\pm f;0)$. Contours of constant $v$, with $v\in[0,2\pi]$, are hyperbolae.  Together these contours form an orthogonal coordinate system.  See Fig. \ref{ellcoords}.  Data is to be interpolated onto the ellipse whose cross section is that of the elliptical cylinder of Fig. \ref{ellipticalcylindergraphic}.  See Fig. \ref{mesh}.

For our work we will need the unit vector $\mbox{\boldmath $\hat{e}$}_u$, the unit vector (outwardly) normal to the surface of the elliptical cylinder.  Write
\begin{eqnarray}
\mbox{\boldmath $r$}&=&
x\mbox{\boldmath $\hat{e}$}_x
+y\mbox{\boldmath $\hat{e}$}_y
+z\mbox{\boldmath $\hat{e}$}_z \nonumber \\
&=&
f\cosh(u)\cos(v)\mbox{\boldmath $\hat{e}$}_x
+f\sinh(u)\sin(v)\mbox{\boldmath $\hat{e}$}_y
+z\mbox{\boldmath $\hat{e}$}_z.\nonumber\\
\label{rhat}
\end{eqnarray}
Then, by definition, we have the result
\begin{eqnarray}
\mbox{\boldmath $\hat{e}$}_u&=&
(\partial \mbox{\boldmath $r$}/\partial u)/||(\partial \mbox{\boldmath $r$}/\partial u)|| \nonumber \\
&=&\frac{\sinh(u)\cos(v)\mbox{\boldmath $\hat{e}$}_x
+\cosh(u)\sin(v)\mbox{\boldmath $\hat{e}$}_y}{[\cosh^2(u)-\cos^2(v)]^{1/2}}.  \label{ehat}
\end{eqnarray}

It is also convenient to employ the complex
variables
\begin{subequations}
\begin{equation}
\zeta=x+iy,  \label{zeta}
\end{equation}
and
\begin{equation}
w=u+iv.  \label{w}
\end{equation}
\end{subequations}
In these variables, the relations (\ref{ellcoord}) can be written in the more compact form
\begin{equation}
\zeta=f\cosh(w). \label{transformation}
\end{equation}
Form differentials of both sides of (\ref{transformation}).  Doing so gives the result
\begin{equation}
dx+idy=f \sinh(w)(du+idv)  \label{differential}
\end{equation}
and the complex conjugate result
\begin{equation}
dx-idy=f \sinh(\bar{w})(du-idv).  \label{differentialc}
\end{equation}
Now form the product of (\ref{differential}) and (\ref{differentialc}) to get the transverse line-element relation
\begin{eqnarray}
ds^2_\bot&=&dx^2+dy^2=f^2\sinh(u+iv)\sinh(u-iv)(du^2+dv^2) \nonumber\\
&=&f^2[\cosh^2(u)-\cos^2(v)](du^2+dv^2).
\end{eqnarray}
From this relation we infer the results
\begin{subequations}
\begin{align}
B_u&=
\mbox{\boldmath $\hat{e}$}_u\cdot\mbox{\boldmath $B$}=
(\nabla\psi)_u  \notag \\
&=
(1/f)[\cosh^2(u)-\cos^2(v)]^{-1/2}({\partial\psi}/{\partial u}),  \label{Bu}
\end{align}
\begin{align}
\nabla^2\psi&=
(1/f^2)[\cosh^2(u)-\cos^2(v)]^{-1}
[(\partial_u)^2+(\partial_v)^2]\psi \nonumber \\
&+(\partial_z)^2\psi.  \label{ellLaplace}
\end{align}
\end{subequations}

\subsubsection{\label{sec:surface:ell:mathieu}Mathieu Equations}

Let us seek to construct harmonic functions of the form
\begin{equation}
\psi\sim P(u)Q(v)\exp(ikz)  \label{ansatz}
\end{equation}
where the functions $P$ and $Q$ are yet to be determined.  Employing the Ansatz (\ref{ansatz}) in Laplace's equation and use of (\ref{ellLaplace}) yields the requirement
\begin{align}
[(\partial_u)^2+(\partial_v)^2]&[P(u)Q(v)]= \nonumber \\
&k^2f^2[\cosh^2(u)-\cos^2(v)]P(u)Q(v).  \label{ellderiv}
\end{align}
We also observe that there is the trigonometric identity
\begin{equation}
\cosh^2(u)-\cos^2(v)=(1/2)[\cosh(2u)-\cos(2v)]
\end{equation}
so that the requirement (\ref{ellderiv}) can be rewritten in the form
\begin{align}
[(\partial_u)^2+(\partial_v)^2]&[P(u)Q(v)]= \label{ellderiv2} \\
(k^2f^2/4)&[2\cosh(2u)-2\cos(2v)]P(u)Q(v).  \nonumber
\end{align}
Upon dividing both sides by $PQ$, (\ref{ellderiv2}) becomes
\begin{align}
(1/P)(\partial_u)^2P&+(1/Q)(\partial_v)^2Q= \nonumber \\
&(k^2f^2/4)[2\cosh(2u)-2\cos(2v)],
\end{align}
from which it follows that
\begin{align}
(1/P)(\partial_u)^2P&-(k^2f^2/4)[2\cosh(2u)]= \\
&-(1/Q)(\partial_v)^2Q-(k^2f^2/4)[2\cos(2v)]. \nonumber
\end{align}
Therefore, there is a common separation constant $a$ such that 
\begin{subequations}
\begin{equation}
(1/P)(\partial_u)^2P-(k^2f^2/4)[2\cosh(2u)]=a
\end{equation}
and
\begin{equation}
-(1/Q)(\partial_v)^2Q-(k^2f^2/4)[2\cos(2v)]=a.
\end{equation}
\end{subequations}
Correspondingly, $P$ and $Q$ must satisfy the ordinary linear differential equations
\begin{subequations}
\begin{equation}
d^2P/du^2-[a-2q\cosh(2u)]P=0,  \label{modMatheq}
\end{equation}
\begin{equation}
d^2Q/dv^2+[a-2q\cos(2v)]Q=0,  \label{ordMatheq}
\end{equation}
where \begin{equation}
q=-k^2f^2/4. \label{qdef}
\end{equation}
\end{subequations}

Equation (\ref{ordMatheq}) for $Q$ is called the {\em Mathieu} equation, and Equation (\ref{modMatheq}) for $P$ is called the {\em modified Mathieu} equation.  For our purposes, we will need solutions $Q(v)$ of (\ref{ordMatheq}) that are periodic with period $2\pi$.  Such solutions exist only for certain {\it characteristic values} of the separation constant $a$.  These values are denoted $a_n(q)$ for $n=0,1,2,3,\cdots$ and $b_n(q)$ for $n=1,2,3,\cdots$.  The solutions associated with the separation constants $a=a_n(q)$ are denoted $\operatorname{ce}_n(v,q)$.  They are even functions of $v$ and, in the small $q$ limit, are proportional to the functions $\cos(nv)$.  The solutions associated with the separation constants $a=b_n(q)$ are denoted $\operatorname{se}_n(v,q)$.  They are odd functions of $v$ and, in the small $q$ limit, are proportional to the functions $\sin(nv)$.  The functions $\operatorname{ce}_n(v,q)$ and $\operatorname{se}_n(v,q)$ form a complete orthogonal set over the interval $v\in[0,2\pi]$ and are normalized so that
\begin{subequations}
\begin{align}
&\int_0^{2\pi}dv{\;} {\rm{ce}}_m(v,q) {\;}{\rm{ce}}_n(v,q)=\pi \delta_{mn}, \\
&\int_0^{2\pi}dv{\;} {\rm{se}}_m(v,q) {\;}{\rm{se}}_n(v,q)=\pi \delta_{mn}, \\
&\int_0^{2\pi}dv{\;} {\rm{ce}}_m(v,q) {\;}{\rm{se}}_n(v,q)=0.
\end{align}
\end{subequations}

With regard to the solutions of the modified Mathieu equation, note that (\ref{ordMatheq}) is transformed into (\ref{modMatheq}) under $v\rightarrow iu$.  As a result, corresponding (real-valued) solutions to (\ref{modMatheq}) are defined by $\operatorname{Ce}_n(u,q)=\operatorname{ce}_n(iu,q)$ and $\operatorname{Se}_n(u,q)=-i\operatorname{se}_m(iu,q)$.  We refer the reader to \cite{Chad, Dragt2} and \cite{McLachlan} for a detailed treatment of the Mathieu functions and their properties.  

\subsubsection{\label{sec:surface:ellgrad}Elliptic Cylinder Harmonic Expansion and On-Axis Gradients}

The stage is now set to describe the expansion of any harmonic function $\psi$ in terms of Mathieu functions.  The general harmonic function that is analytic in $x$ and $y$ near the origin can be written in the coordinates (\ref{ellcoord}) in the form
\begin{align}
\psi(u,v,z)&=
\sum_{n=0}^\infty\int_{-\infty}^\infty dk{\;}c_n(k)e^{ikz}{\rm{Ce}}_{n}(u,q){\;}{\rm{ce}}_{n}(v,q)\nonumber \\
&+\sum_{n=1}^\infty\int_{-\infty}^\infty dk{\;}s_n(k)e^{ikz}{\rm{Se}}_{n}(u,q){\;}{\rm{se}}_{n}(v,q)\label{ellharmonic} 
\end{align}
where the functions $c_n(k)$ and $s_n(k)$ are arbitrary.  We will call (\ref{ellharmonic}) an {\em elliptic cylinder harmonic} expansion.

To exploit this expansion, suppose the magnetic field $\mbox{\boldmath $B$}(x,y,z)$ is interpolated onto the surface $u=U$ of an elliptic cylinder using values at the grid points near the surface. See Fig. \ref{mesh}. Let us employ the notation $\mbox{\boldmath $B$}(x,y,z)=\mbox{\boldmath $B$}(u,v,z)$ so that the magnetic field on the surface can be written as $\mbox{\boldmath $B$}(U,v,z)$.  Next, from the values on the surface, compute $B_u(U,v,z)$, the component of $\mbox{\boldmath $B$}(x,y,z)$ {\em normal} to the surface.  Our aim will be to determine the on-axis gradients from a knowledge of $B_u(U,v,z)$.  At this point we note that the functions $\exp(ikz){\rm {se}}_n(v,q)$ and $\exp(ikz){\rm {ce}}_n(v,q)$ form a complete set over the surface of the elliptical cylinder.

Let us begin by solving (\ref{Bu}) for $(\partial\psi/\partial u)$.  We find, using (\ref{ehat}), the result,
\begin{eqnarray}
(\partial\psi/\partial u)&=&
f[\cosh^2(u)-\cos^2(v)]^{1/2}B_u \label{dpsidu} \\
&=&
f(\sinh u \cos v)B_x+f(\cosh u \sin v)B_y.  \nonumber
\end{eqnarray}
We see that the right side of (\ref{dpsidu}) is a  well-behaved function $F(u,v,z)$ whose values are known for $u=U$,
\begin{align}
F(U,v,z)&=f(\sinh U \cos v)B_x(U,v,z) \nonumber \\
&+f(\cosh U \sin v)B_y(U,v,z).  \label{Fu}
\end{align}

Moreover, using the representation (\ref{ellharmonic}) in (\ref{dpsidu}) and (\ref{Fu}), we may also write
\begin{align}
F(U,v,z)&=
\sum_{n=0}^\infty\int_{-\infty}^\infty dk{\;}c_n(k)e^{ikz}{\rm{Ce}}_{n}^\prime(U,q){\;}{\rm{ce}}_{n}(v,q)\nonumber \\
&+\sum_{n=1}^\infty\int_{-\infty}^\infty dk{\;}s_n(k)e^{ikz}{\rm{Se}}_{n}^\prime(U,q){\;}{\rm{se}}_{n}(v,q).\label{Fu2}
\end{align}
Next multiply both sides of (\ref{Fu2}) by $\exp(-ik^\prime z)$ and integrate over $z$.  So doing gives the result
\begin{align}
\frac{1}{2\pi}\int_{-\infty}^\infty dz{\;} e^{-ik z}F(U,v,z)&=\sum_{n=0}^\infty c_n(k){\rm{Ce}}_{n}^\prime(U,q){\;}{\rm{ce}}_{n}(v,q)\nonumber \\
&
+\sum_{n=1}^\infty s_n(k){\rm{Se}}_{n}^\prime(U,q){\;}{\rm{se}}_{n}(v,q). 
\end{align}
Now, employ the orthogonality properties of the Mathieu functions to obtain the relations
\begin{subequations}\label{cCsS}
\begin{align}
c_r(k)&{\rm{Ce}}_{r}^\prime(U,q)= \nonumber \\
&\frac{1}{2\pi^2}\int_0^{2\pi}dv{\;} {\rm{ce}}_{r}(v,q)\int_{-\infty}^\infty dz{\;} e^{-ik z}F(U,v,z), \\
s_r(k)&{\rm{Se}}_{r}^\prime(U,q)= \nonumber \\
&\frac{1}{2\pi^2}\int_0^{2\pi}dv{\;} {\rm{se}}_{r}(v,q)\int_{-\infty}^\infty dz{\;} e^{-ik z}F(U,v,z).
\end{align}
\end{subequations}
In view of (\ref{cCsS}), define the function $\tilde{F}(v,k)$ by the rule
\begin{equation}
\tilde{F}(v,k)=\frac{1}{2\pi}\int_{-\infty}^\infty dz{\;} e^{-ik z}F(U,v,z),
\end{equation}
and define functions $\tilde{\tilde{F}}^c_r(k)$ and $\tilde{\tilde{F}}^s_r(k)$ by the rules
\begin{subequations}\label{intermell}
\begin{align}
\tilde{\tilde{F}}^c_r(k)&=\frac{1}{\pi}\int_0^{2\pi}dv {\;}{\rm{ce}}_{r}(v,q)\tilde{F}(v,k)\nonumber\\
&=\frac{1}{2\pi^2}\int_0^{2\pi}dv{\;} {\rm{ce}}_{r}(v,q)\int_{-\infty}^\infty dz{\;} e^{-ik z}F(U,v,z), \\
\tilde{\tilde{F}}^s_r(k)&=\frac{1}{\pi}\int_0^{2\pi}dv {\;}{\rm{se}}_{r}(v,q)\tilde{F}(v,k)\nonumber\\
&=\frac{1}{2\pi^2}\int_0^{2\pi}dv{\;} {\rm{se}}_{r}(v,q)\int_{-\infty}^\infty dz{\;} e^{-ik z}F(U,v,z).
\end{align}
\end{subequations}
Note the similarity to (\ref{intermcyl}), where $\cos(r\phi)$ and $\sin(r\phi)$ are replaced by ${\rm{ce}}_{r}(v,q)$ and ${\rm{se}}_{r}(v,q)$.  We will call the functions $\tilde{\tilde{F}}^\alpha_r(k)$ {\em Mathieu coefficient} functions in analogy to the Fourier coefficients that arise in Fourier analysis.

With these definitions, the relations (\ref{cCsS}) can be rewritten in the form
\begin{equation}
c_r(k)=\frac{\tilde{\tilde{F}}^c_r(k)}{{\rm{Ce}}_{r}^\prime(U,q)},\quad\quad
s_r(k)=\frac{\tilde{\tilde{F}}^s_r(k)}{{\rm{Se}}_{r}^\prime(U,q)}.  \label{cs}
\end{equation}
Finally, employ (\ref{cs}) in (\ref{ellharmonic}).  So doing gives the result
\begin{align}
\psi&(x,y,z)= \nonumber \\
&\sum_{r=0}^\infty\int_{-\infty}^\infty dk{\;}e^{ikz}[\tilde{\tilde{F}}^c_r(k)/{\rm{Ce}}_{r}^\prime(U,q)]{\rm{Ce}}_{r}(u,q){\;}{\rm{ce}}_{r}(v,q)\nonumber \\
+&\sum_{r=1}^\infty\int_{-\infty}^\infty dk{\;}e^{ikz}[\tilde{\tilde{F}}^s_r(k)/{\rm{Se}}_{r}^\prime(U,q)]{\rm{Se}}_{r}(u,q){\;}{\rm{se}}_{r}(v,q).\label{ellharmonic2}
\end{align}
We have obtained an elliptical cylinder harmonic expansion for $\psi$ in terms of surface field data.

Of course, what we really want are the on-axis gradients.  Again, once these gradients are known, we may use (\ref{psieq}) and (\ref{Atran}) to compute the associated scalar and vector potentials.  The gradients can  be found by employing two remarkable {\em connections} (identities) between elliptic and circular cylinder functions of the form \cite{Chad}
\begin{subequations}
\begin{equation}
{\rm{Ce}}_{r}(u,q){\;}{\rm{ce}}_{r}(v,q)=
\sum_{m=0}^\infty \alpha^r_m(k)I_m(k\rho)\cos(m\phi),
\end{equation}
\begin{equation}
{\rm{Se}}_{r}(u,q){\;}{\rm{se}}_{r}(v,q)=
\sum_{m=1}^\infty \beta^r_m(k)I_m(k\rho)\sin(m\phi).
\end{equation}
\end{subequations}
For further reference, we will call the quantities $\alpha^r_m(k)$ and $\beta^r_m(k)$ {\em Mathieu-Bessel connection coefficients} \cite{Chad, Dragt2}.
Using these results, (\ref{ellharmonic2}) can be rewritten in the form
\begin{align}
\psi&(x,y,z)= \nonumber \\
&\sum_{m=0}^\infty \int_{-\infty}^\infty dk{\;}e^{ikz}I_m(k\rho)\cos(m\phi)\sum_{r=0}^\infty \alpha^r_m(k)[\tilde{\tilde{F}}^c_r(k)/{\rm{Ce}}_{r}^\prime(U,q)] \nonumber \\
+&\sum_{m=1}^\infty \int_{-\infty}^\infty dk{\;}e^{ikz}I_m(k\rho)\sin(m\phi)\sum_{r=1}^\infty \beta^r_m(k)[\tilde{\tilde{F}}^s_r(k)/{\rm{Se}}_{r}^\prime(U,q)]. \label{ellharmonic3}
\end{align}
Upon comparing (\ref{ellharmonic3}) with (\ref{Laplacecylinder}), we conclude that there are the relations
\begin{subequations}\label{ellharmoniccoeff}
\begin{equation}
G_{m,c}(k)=
\sum_{r=0}^\infty \alpha^r_m(k)[\tilde{\tilde{F}}^c_r(k)/{\rm{Ce}}_{r}^\prime(U,q)],
\end{equation}
and
\begin{equation}
G_{m,s}(k)=
\sum_{r=1}^\infty \beta^r_m(k)[\tilde{\tilde{F}}^s_r(k)/{\rm{Se}}_{r}^\prime(U,q)].
\end{equation}
\end{subequations}
Finally, in view of (\ref{CfromG}) and (\ref{ellharmoniccoeff}), we have the desired results
\begin{subequations}\label{ellipticalgradients}
\begin{align}
C^{[n]}_{m,c} (z)&= \nonumber \\
\frac{i^n}{2^{m}m!}&\int_{-\infty}^\infty dk{\;} e^{ikz}k^{n+m} 
\sum_{r=0}^\infty \alpha^r_m(k)[\tilde{\tilde{F}}^c_r(k)/{\rm{Ce}}_{r}^\prime(U,q)],
\end{align}
\begin{align}
C^{[n]}_{m,s} (z)&= \nonumber \\
\frac{i^n}{2^{m}m!}&\int_{-\infty}^\infty dk{\;} e^{ikz}k^{n+m} 
\sum_{r=1}^\infty \beta^r_m(k)[\tilde{\tilde{F}}^s_r(k)/{\rm{Se}}_{r}^\prime(U,q)].
\end{align}
\end{subequations}
We have found expressions for the on-axis gradients in terms of field data (normal component) on the surface of an elliptic cylinder.

\subsection{\label{sec:surface:rect} Use of Field Data on Surface of Rectangular Cylinder}
A similar procedure has been developed for computing the on-axis gradients in terms of field data provided on the surface of a rectangular cylinder.  In this case, each on-axis gradient $C^{[n]}_{m,\alpha}$ may be written as the sum of four contributions,
\begin{equation}
C^{[n]}_{m,\alpha}(z)=\sum_{\beta=T,B,L,R}{\;}^\beta C^{[n]}_{m,\alpha}(z).
\end{equation}
Each contribution is determined by the integration of the normal component of the field against an appropriate kernel over one of the four faces (Top, Bottom, Left, Right) of the rectangular cylinder.  The resulting expressions are quite lengthy, and we therefore refer the reader to \cite{Chad, Dragt2} for further details.  The remainder of this paper will focus on the circular and elliptical cylinder cases.

\section{\label{sec:benchmark} Numerical Benchmarks}
\subsection{\label{sec:benchmark:doublet}Monopole Doublet}
\begin{figure}[hpt]
  \centering
   \resizebox{8.6cm}{!}{  \includegraphics*[width=4in,angle=-90]{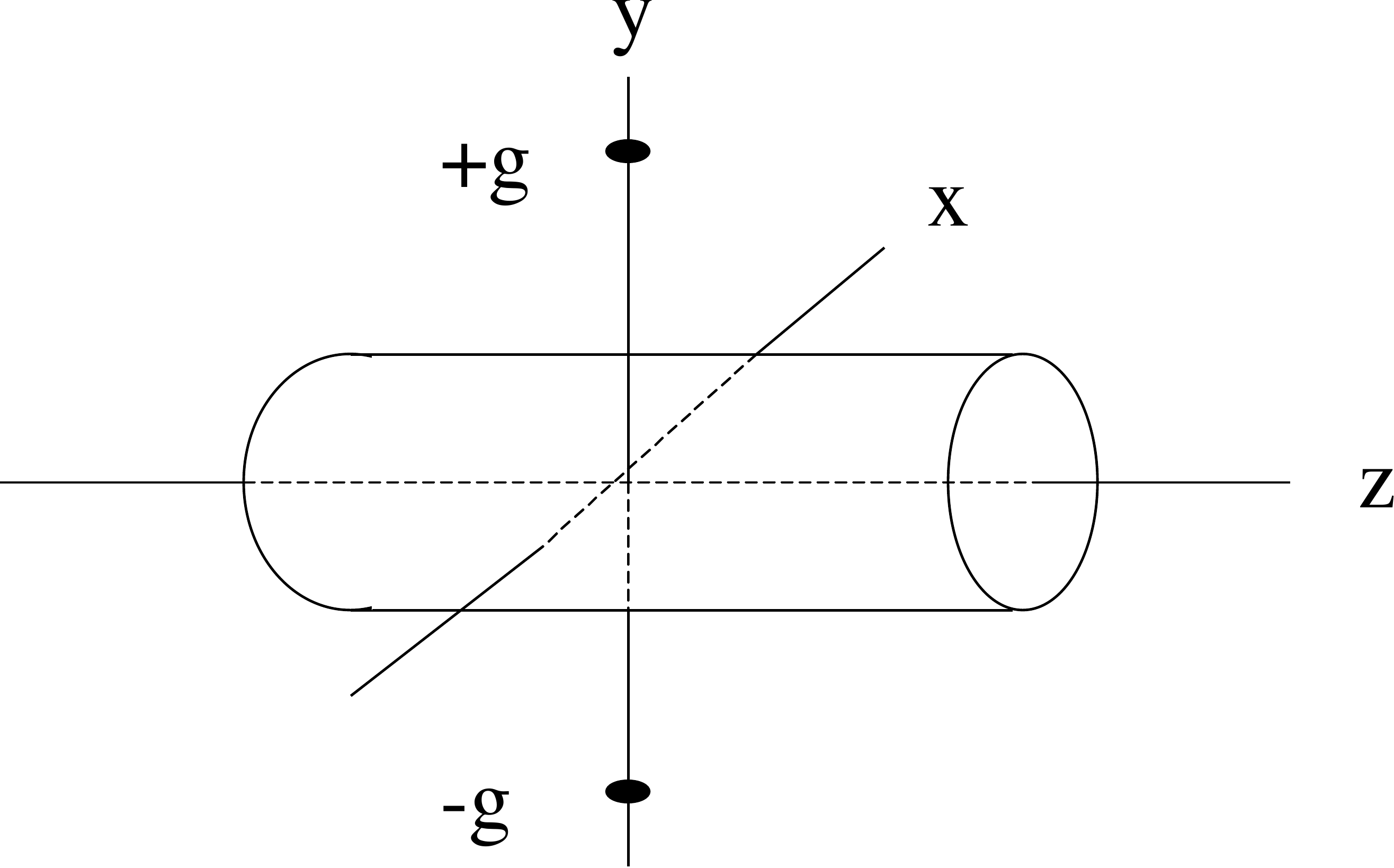}}
  \caption{\label{doubletgraphic} A monopole doublet consisting of two magnetic monopoles of equal and opposite sign placed on the $y$ axis and centered on the origin.   Also shown, for future reference, is a cylinder with circular cross section placed in the interior field.}
\end{figure}

\begin{figure}[hpt]
  \centering
    \resizebox{8.6cm}{!}{ \includegraphics*[width=4in,angle=-90]{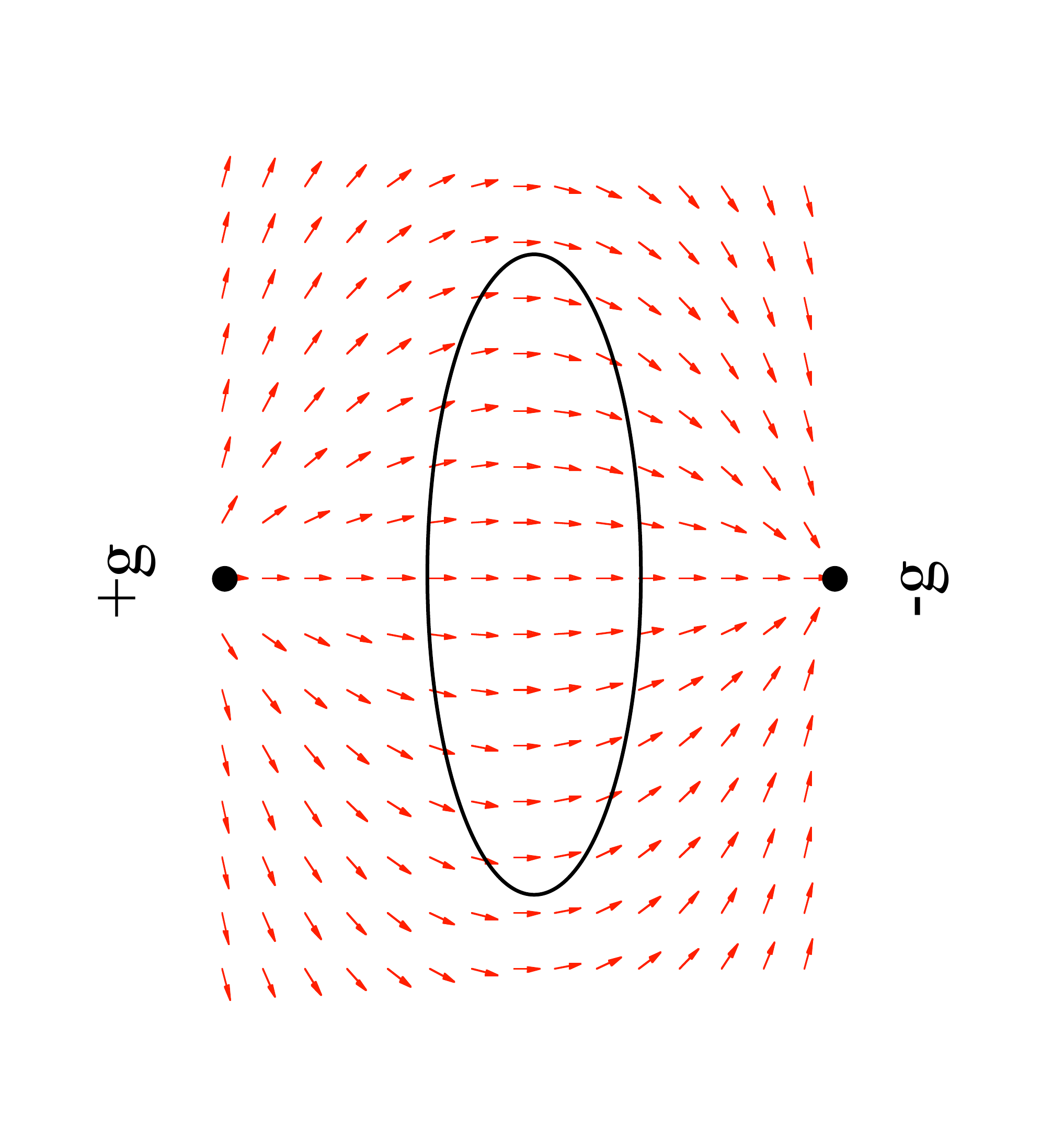}}
  \caption{\label{doubletfield} The interior field of a monopole doublet.  Also shown is an ellipse which will be used in Section \ref{sec:benchmark:ell}.}
\end{figure}

\begin{figure}[hpt]
  \centering
    \resizebox{8.6cm}{!}{ \includegraphics*[height=3.5in,angle=0]{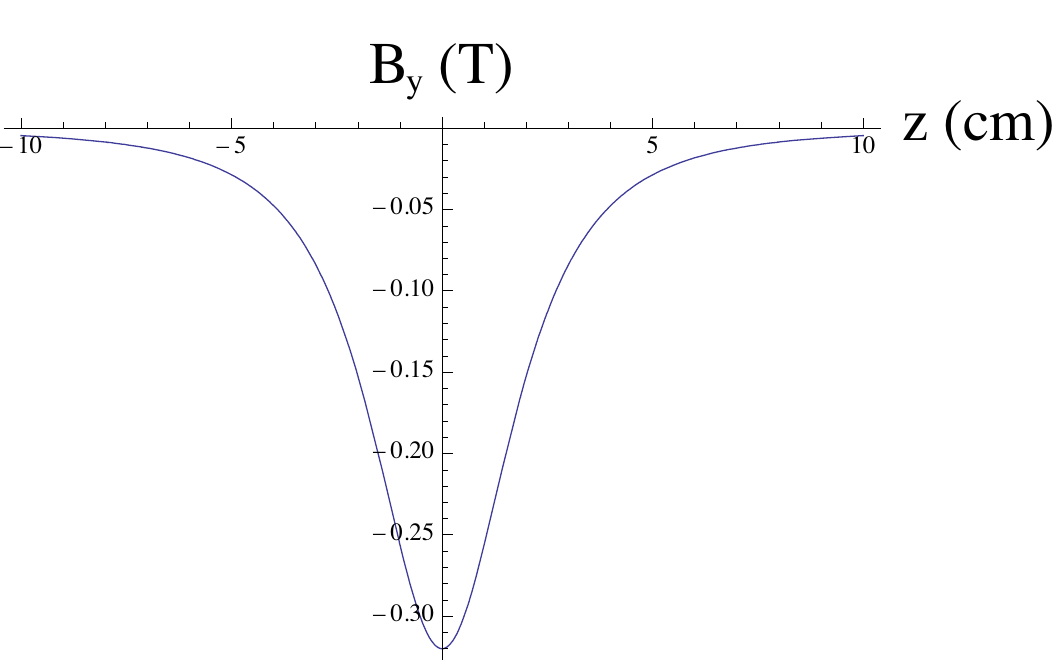}}
  \caption{\label{doubletfieldBy} The on-axis field component $B_y(x=0,y=0,z)$ for the monopole doublet in the case that $a=2.5$ cm and $g=1$ Tesla-$({\rm cm})^2$. The coordinate $z$ is given in centimeters.}
\end{figure}

\begin{figure}[hpt]
  \centering
  \resizebox{8.6cm}{!}{ \includegraphics*[height=3.5in,angle=0]{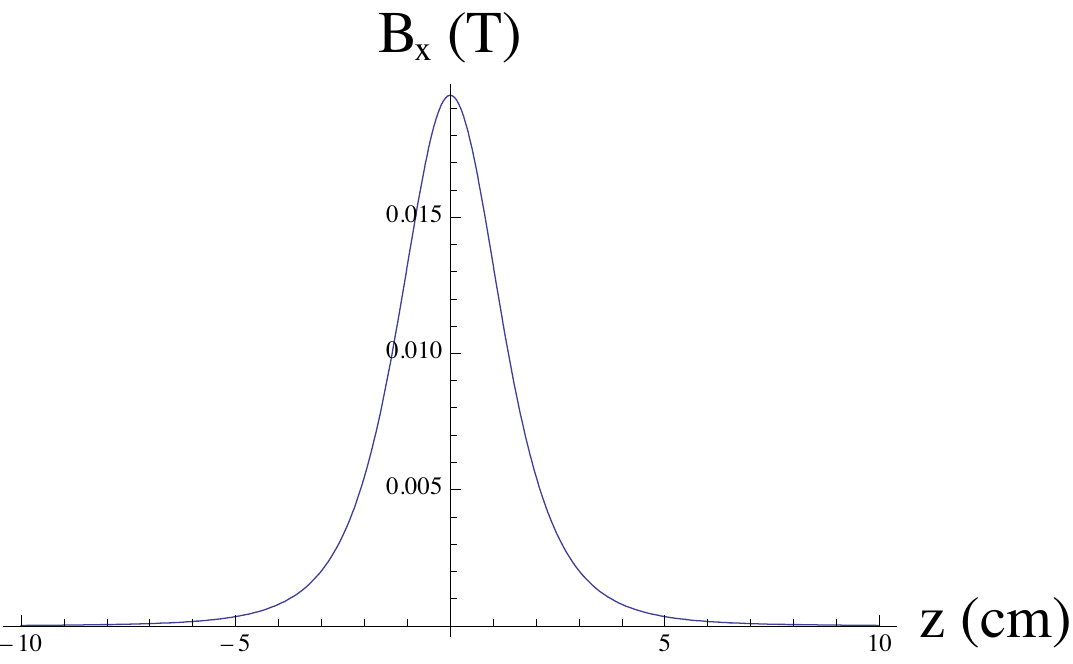}}
  \caption{\label{doubletfieldBx} The field component $B_x$ on the line $\rho=1/2$, $\phi=\pi/4$, $z\in[-\infty,\infty]$ for the monopole doublet in the case that $a=2.5$ cm and $g=1$ Tesla-$({\rm cm})^2$. The coordinate $z$ is given in centimeters.}
\end{figure}

\begin{figure}[hpt]
  \centering
    \resizebox{8.6cm}{!}{ \includegraphics*[height=3.5in,angle=0]{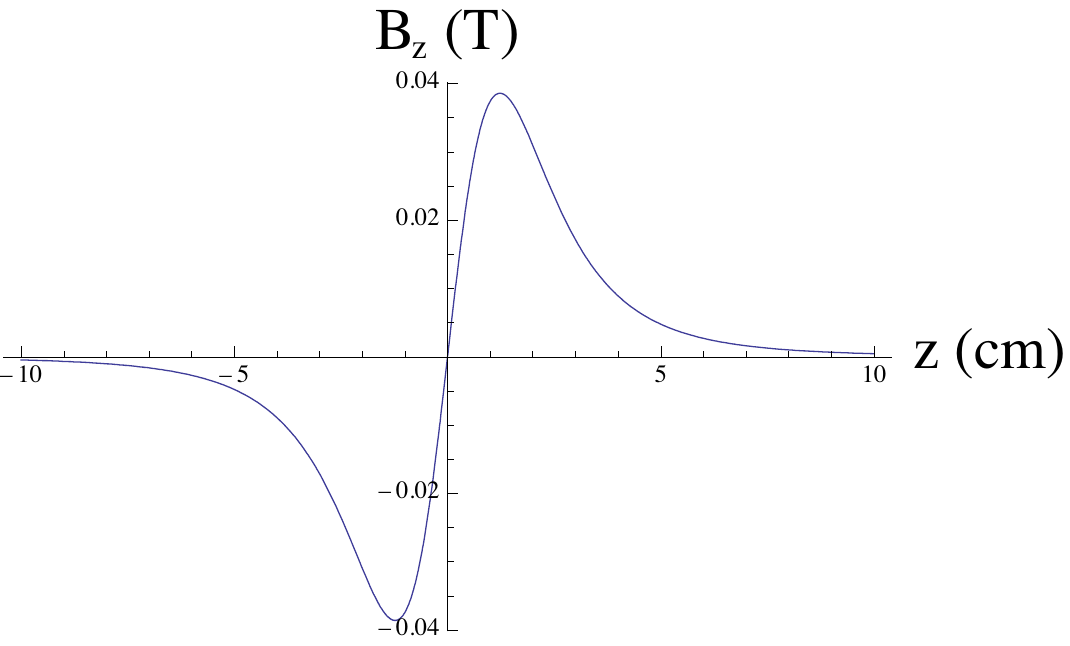}}
  \caption{\label{doubletfieldBz} The field component $B_z$ on the line $\rho=1/2$, $\phi=\pi/4$, $z\in[-\infty,\infty]$ for the monopole doublet in the case that $a=2.5$ cm and $g=1$ Tesla-$({\rm cm})^2$. The coordinate $z$ is given in centimeters.}
\end{figure}

In this section, we develop an exactly-soluble but numerically challenging model field to be used to numerically benchmark the procedures developed in Section III.  Suppose two magnetic monopoles having strengths $\pm g$ are placed at the $(x,y,z)$ locations
\begin{align}
r^+&=(0,a,0), \nonumber \\
r^-&=(0,-a,0).
\end{align}
See Fig. \ref{doubletgraphic}, which also shows a circular cylinder with radius $R$ (the surface $\rho=R$).  These monopoles generate a scalar potential $\psi(x,y,z)$ described by the relation
\begin{align}
\psi(x,y,z)&= \nonumber \\
-g[x^2+&(y-a)^2+z^2]^{-1/2}+g[x^2+(y+a)^2+z^2]^{-1/2}\nonumber \\
=\psi_+&(x,y,z)+\psi_-(x,y,z).
\end{align}
Correspondingly, they produce a magnetic field $\mbox{\boldmath $B$} = \nabla \psi$ having the components
\begin{subequations}\label{doubletBeq}
\begin{equation}
B_x=gx[x^2+(y-a)^2+z^2]^{-3/2}-gx[x^2+(y+a)^2+z^2]^{-3/2},
\end{equation}
\begin{align}
B_y=g(y-a)\{[x^2+&(y-a)^2+z^2]^{-3/2} \nonumber \\
&-g(y+a)[x^2+(y+a)^2+z^2]^{-3/2}\},
\end{align}
\begin{equation}
B_z=gz[x^2+(y-a)^2+z^2]^{-3/2}-gz[x^2+(y+a)^2+z^2]^{-3/2}.
\end{equation}
\end{subequations}
This field is sketched in Fig. \ref{doubletfield}.  To provide further insight, Fig. \ref{doubletfieldBy} shows the on-axis field component $B_y(x=0,y=0,z)$, and Figs. \ref{doubletfieldBx} and \ref{doubletfieldBz} show the off-axis field components $B_x(\rho=1/2,\phi=\pi/4,z)$ and $B_z(\rho=1/2,\phi=\pi/4,z)$.

Due to the symmetries of the field, the functions $C^{[n]}_{m,c}$ vanish for all $m$.  Furthermore, $C^{[n]}_{m,s}=0$ for $m$ even.  It can be shown that the nonvanishing on-axis gradients are given for $m$ odd by \cite{Chad}:
\begin{subequations}\label{monopolegradientexpr}
\begin{equation}
C^{[0]}_{m,s}(z)=(-1)^{(m-1)/2}\frac{g}{a^{m+1}}\frac{(2m)!}{2^{2m-2}(m!)^2}
\beta^{2m+1}(z) 
\end{equation}
where
\begin{equation}
\beta(z)=\frac{a}{\sqrt{z^2+a^2}}.
\end{equation}
\end{subequations}
For fixed $z$, the domain of convergence for the polynomial series (\ref{psieq}-\ref{Atran}) representing the field in terms of the functions (\ref{monopolegradientexpr}) is given by the condition $\sqrt{x^2+y^2}<\sqrt{z^2+a^2}$.  In particular, the domain of convergence is a region of circular cross-section whose radius increases as we move longitudinally away from the location of the monopoles at $z=0$.  

Suppose we wish to calculate a transfer map through $7^{\rm th}$ order.  Then, as shown in the Appendix, to do so requires knowledge of the $C^{[n]}_{m,\alpha}(z)$ with $(m+n)\le7$ when $m=0$ or $m$ is odd, and knowledge of the $C^{[n]}_{m,\alpha}(z)$ with $(m+n)\le8$ when $m$ is even.

Graphs of a selected few of these functions, for the monopole doublet in the case that $a=2.5$ cm and $g=1$ Tesla-$({\rm cm})^2$, are shown in Figs. \ref{CircC10} through \ref{CircC70}. In these plots $z$ has units of centimeters.  Evidently the $C^{[0]}_{m,s}$ become ever more highly peaked with increasing $m$.  Fortunately, when working through some fixed degree, we need fewer derivatives with increasing $m$. Note that we expect that the function $C^{[n]}_{m,s}(z)$ should have $n$ zeroes.  This is indeed the case.

\subsection{\label{sec:benchmark:circ}Circular Cylinder Results}
The procedure discussed in Section \ref{sec:surface:circ} has been benchmarked using the field of a monopole doublet in the case that $a=2.5$ cm and $g=1$ Tesla-$({\rm cm})^2$.
We set up a regular grid in $x$, $y$, $z$ space, where we let each variable range over the intervals $x\in [-4.4,4.4]$ with spacing $h_x=0.1$, $y\in [-2.4,2.4]$ with $h_y=0.1$, and $z\in [-300,300]$ with $h_z=0.125$ (in units of cm).  The known values of the three components of the field are computed using (\ref{doubletBeq}) at each grid point.  Consider a cylinder of radius $R=2$ cm and length $600$ cm.  We use bicubic interpolation to interpolate $\mbox{\boldmath $B$}$ at these grid points onto $49$ selected angular points on the cylinder, for each of the 4801 selected values of $z$.  The angular integration in (\ref{intermcyl}) is performed using a Riemann sum with $N=49$.  (This is necessary to ensure sufficient convergence of the angular integrals to within $10^{-4}$.)  We evaluate the Fourier transform at 401 values of $k$ in the range $[-K_c,K_c]$ with $K_c=20$, using a spline-based Fourier transform algorithm \cite{Chad, Dragt2}.  We use these same points in $k$ space to evaluate the inverse Fourier transform, providing a set of numerically determined functions $C_{m,\alpha}^{[n]}(z)$.

\begin{figure}[hpt]
  \centering
   \resizebox{8.6cm}{!}{ \includegraphics*[height=3.5in,angle=0]{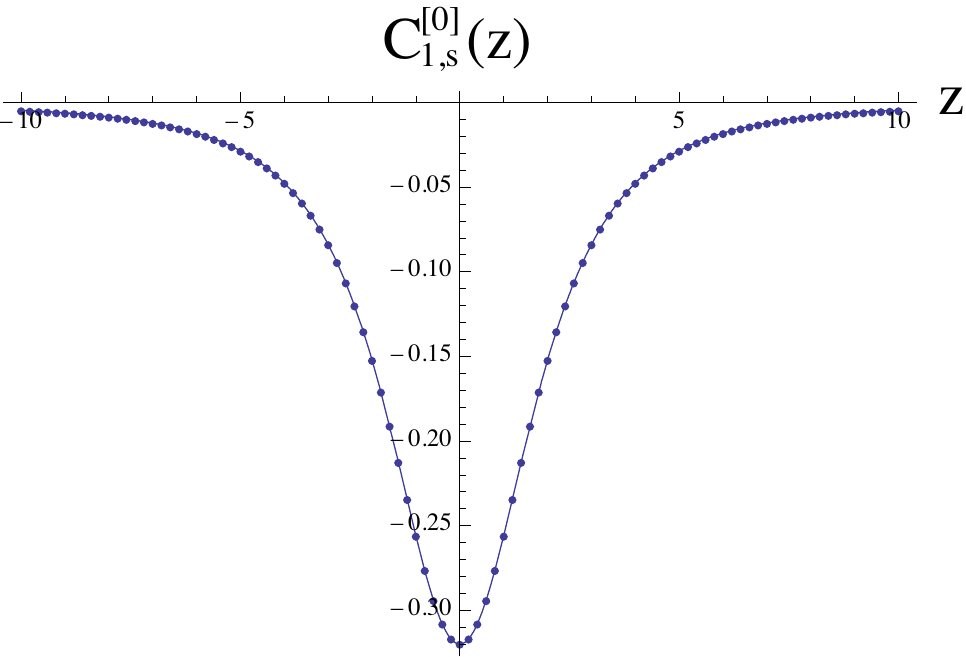}}
  \caption{\label{CircC10} Exact and numerical results for $C^{[0]}_{1,s}(z)$.  Exact results are shown as a solid line, and numerical results are shown as dots.}
\end{figure}

\begin{figure}[hpt]
  \centering
    \resizebox{8.6cm}{!}{\includegraphics*[height=3.5in,angle=0]{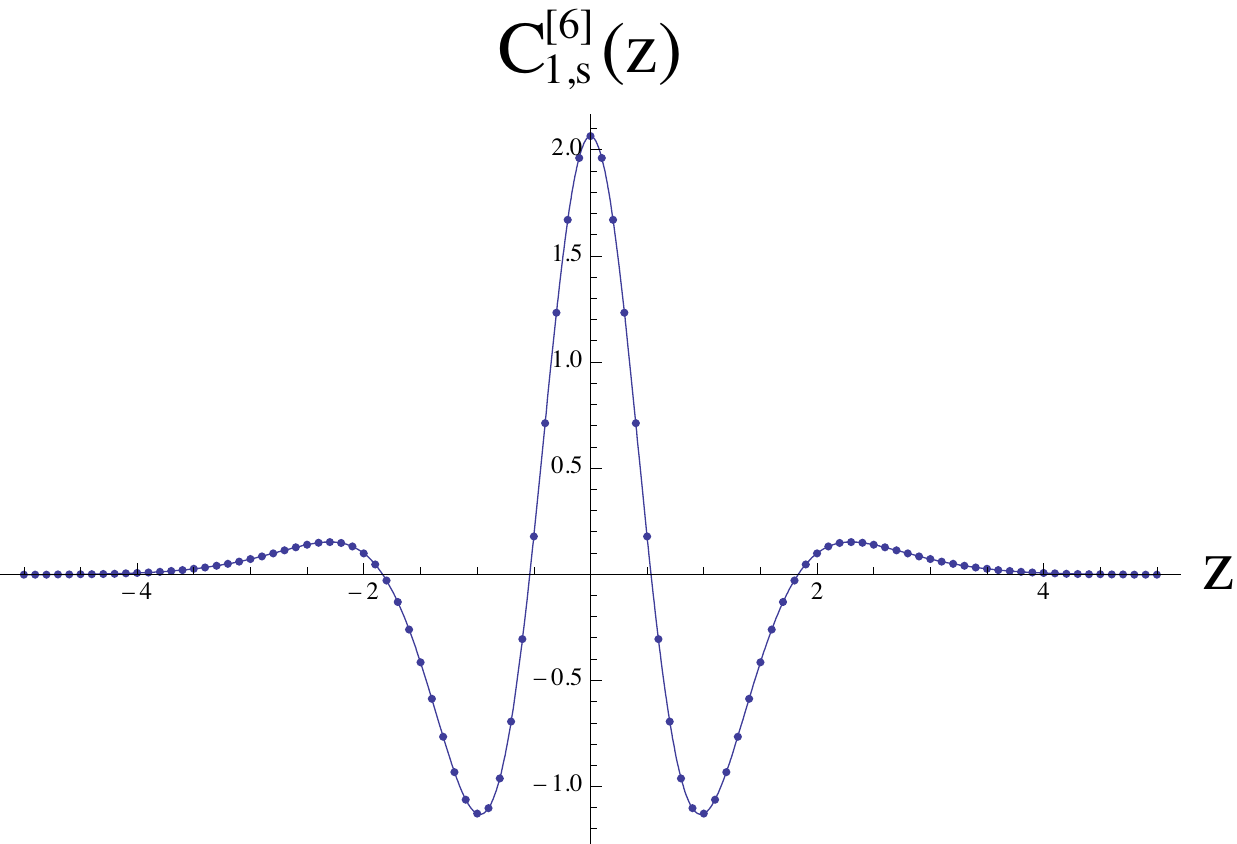}}
  \caption{\label{CircC16} Exact and numerical results for $C^{[6]}_{1,s}(z)$.  Exact results are shown as a solid line, and numerical results are shown as dots.}
\end{figure}

\begin{figure}[hpt]
  \centering
 \resizebox{8.6cm}{!}{\includegraphics*[height=3.5in,angle=0]{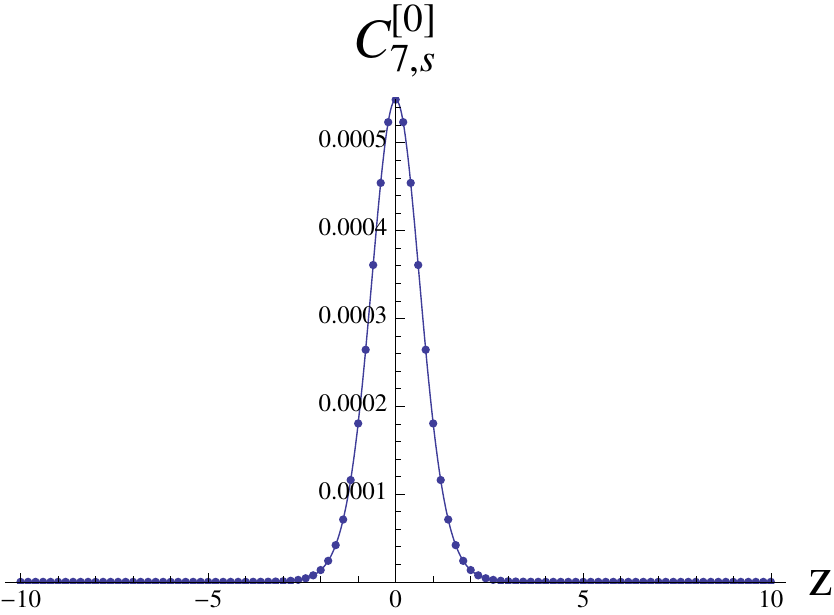}}
  \caption{\label{CircC70} Exact and numerical results for $C^{[0]}_{7,s}(z)$.  Exact results are shown as a solid line, and numerical results are shown as dots.}
\end{figure}

\begin{figure}[hpt]
  \centering
   \resizebox{9.0cm}{!}{ \includegraphics*[height=3.5in,angle=0]{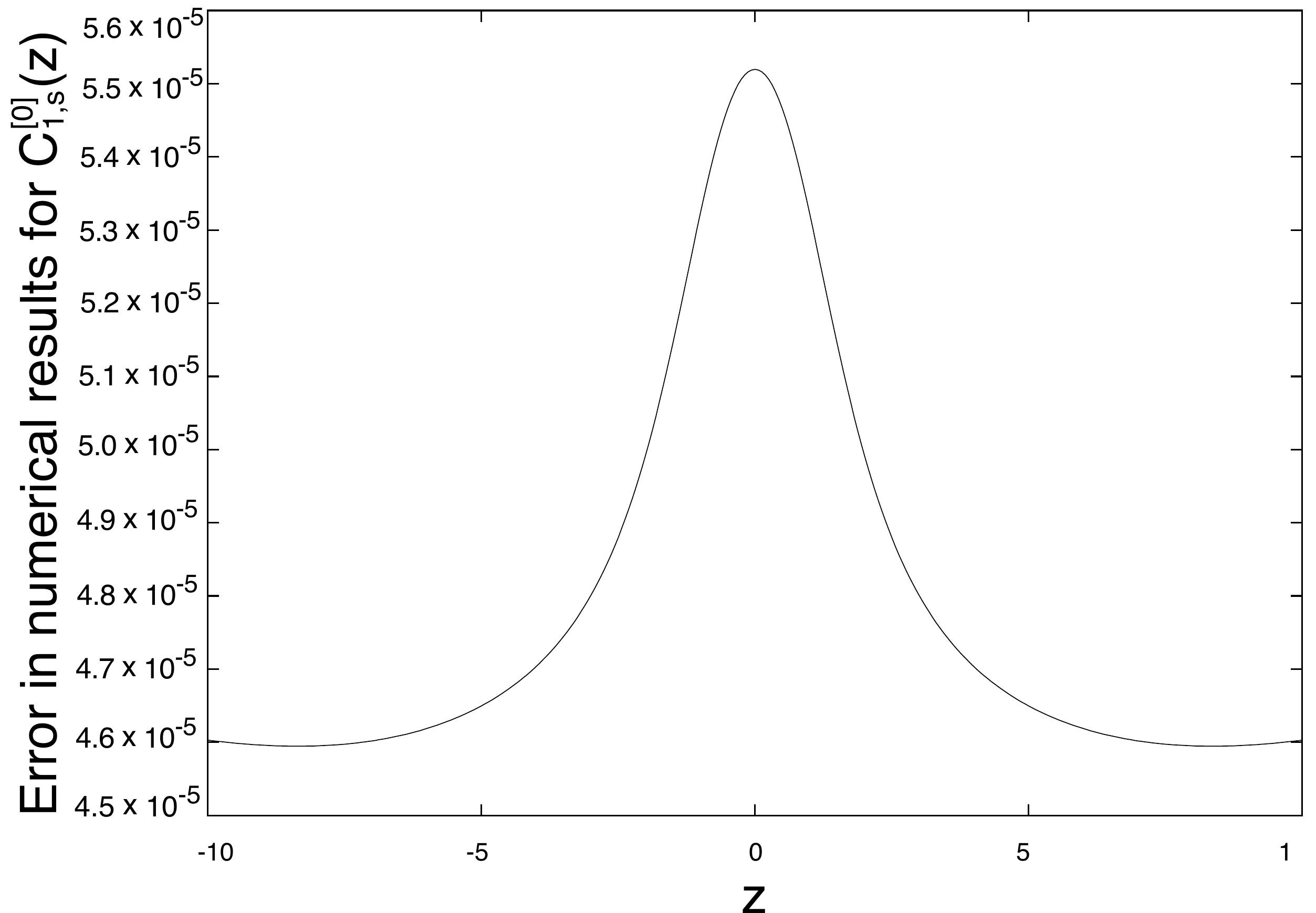}}
  \caption{\label{CircC10error} Difference between exact and numerical results for $C^{[0]}_{1,s}(z)$.}
\end{figure}

A comparison between the exact on-axis gradients (\ref{monopolegradientexpr}) and those obtained from grid data is provided in Figs. \ref{CircC10}-\ref{CircC70} for the functions $C_{1,s}^{[0]}$, $C_{1,s}^{[6]}$, and $C_{7,s}^{[0]}$.  Evidently the agreement is excellent.  Fig. \ref{CircC10error} illustrates the difference between the on-axis gradient $C^{[0]}_{1,s}$ as obtained from grid data and the exact on-axis gradient obtained from (\ref{monopolegradientexpr}) with $m=1$.  The maximum error attained relative to peak is $1.7\times 10^{-4}$.  Further detailed study shows that numerical results agree with exact results to within relative errors less than a few parts in $10^4$ for all the relevant $C_{m,\alpha}^{[n]}(z)$.  By using exact data on the cylinder rather than interpolating off the grid onto the cylinder, we have also verified that the error due to interpolation onto the cylinder is comparable to that produced by numerical integration\cite{Dragt2}.  Finally, all these small errors can be further reduced with the aid of a finer grid \cite{Chad, Dragt2}.

\subsection{\label{sec:benchmark:ell}Elliptical Cylinder Results}
The procedure discussed in Section \ref{sec:surface:ell} has been benchmarked using grid values identitical to those described in the previous section.
Consider an elliptical cylinder of semiminor axis of $y_{\rm max}=2$ cm and semimajor axis $x_{\rm max}=4$ cm.  In this case, we evaluate the angular integrals (\ref{intermell}) using a Riemann sum with $N=120$.  (This is necessary to ensure sufficient convergence of the angular integrals to within $10^{-4}$.)  Doing so requires interpolation off the grid onto the elliptical cylinder at 120 angular points for each of the 4801 selected values of $z$.  The sums in (\ref{ellipticalgradients}) are truncated beyond $r=r_{max}$, where $r_{max}$ varies from 11 to 29 as necessary to achieve a tolerance of 1 part in $10^4$.  We evaluate the Fourier transform at 401 values of $k$ in the range $[-K_c,K_c]$ with $K_c=20$, using a spline-based Fourier transform algorithm.  We use these same points in $k$ space to evaluate the inverse Fourier transform, providing a set of numerically determined functions $C_{m,\alpha}^{[n]}(z)$.

Results for the functions  $C_{m,\alpha}^{[n]}(z)$ are similar to those found in the circular cylinder case \cite{Chad, Dragt2}, and have comparable accuracy.  In the following section, however, we illustrate that functions obtained using an elliptical cylinder are significantly more robust against numerical noise in the original grid values.

\section{\label{sec:smoothing}Smoothing}
In this section, we investigate the smoothing of numerical noise that results from the use of the surface fitting techniques described in Section \ref{sec:surface}.  Consider computing the on-axis gradients $C_{m,\alpha}^{[n]}$ from a grid of numerical field values $\mbox{\boldmath $B$}$ using a circular or elliptical cylinder, as described in the previous sections.  Observe that (\ref{circulargradients}) and (\ref{ellipticalgradients})
are linear in each of the values $B_x$ and $B_y$ at the grid points.  The inclusion of  additive numerical noise $\Delta\mbox{\boldmath $B$}$ at each grid point therefore results in on-axis gradients of the form $C_{m,\alpha}^{[n]}+\Delta C_{m,\alpha}^{[n]}$, where the contribution $\Delta C_{m,\alpha}^{[n]}$ is determined by the values $\Delta\mbox{\boldmath $B$}$ according to the procedures of Section \ref{sec:surface}.  Note that only the field values $\Delta B_x$ and $\Delta B_y$ at grid points near the surface of the circular cylinder affect the functions $\Delta C_{m,\alpha}^{[n]}$.

To examine the effect of additive noise, we generate a random noise field $\Delta\mbox{\boldmath $B$}$ whose components are proportional, at the 1\% level, to the strength of the monopole-doublet on-axis vertical field. 
Consider the grid $[-4.4,4.4]\times[-2.4,2.4]\times[-300,300]$ cm used in Section \ref{sec:benchmark} for fitting the field of the monopole doublet, with mesh points indexed by $j=1,\cdots N$.  Let $B_y(0,0,z)$ denote the value of the monopole-doublet on-axis vertical field at longitudinal location $z$, as determined from (\ref{doubletBeq}).  At each point $(x_j,y_j,z_j)$ we set 
\begin{subequations}\label{noisefield}
\begin{align}
\Delta B_x(x_j,y_j,z_j)=\epsilon B_y(0,0,z_j)\delta_x(j),\\
\Delta B_y(x_j,y_j,z_j)=\epsilon B_y(0,0,z_j)\delta_y(j).
\end{align}
\end{subequations}
Here the $\delta_x(j)$ and $\delta_y(j)$ are uniformly distributed random variables taking values in the interval $[-1,1]$, and $\epsilon=0.01$.  After interpolating these values onto the surface of a circular cylinder with $R=2$ cm and $z\in[-300,300]$ cm, we use the procedure described in Section IIIA to compute the on-axis gradients (\ref{circulargradients}).

\begin{figure}[hpt]
  \centering
  \resizebox{8.6cm}{!}{ \includegraphics*{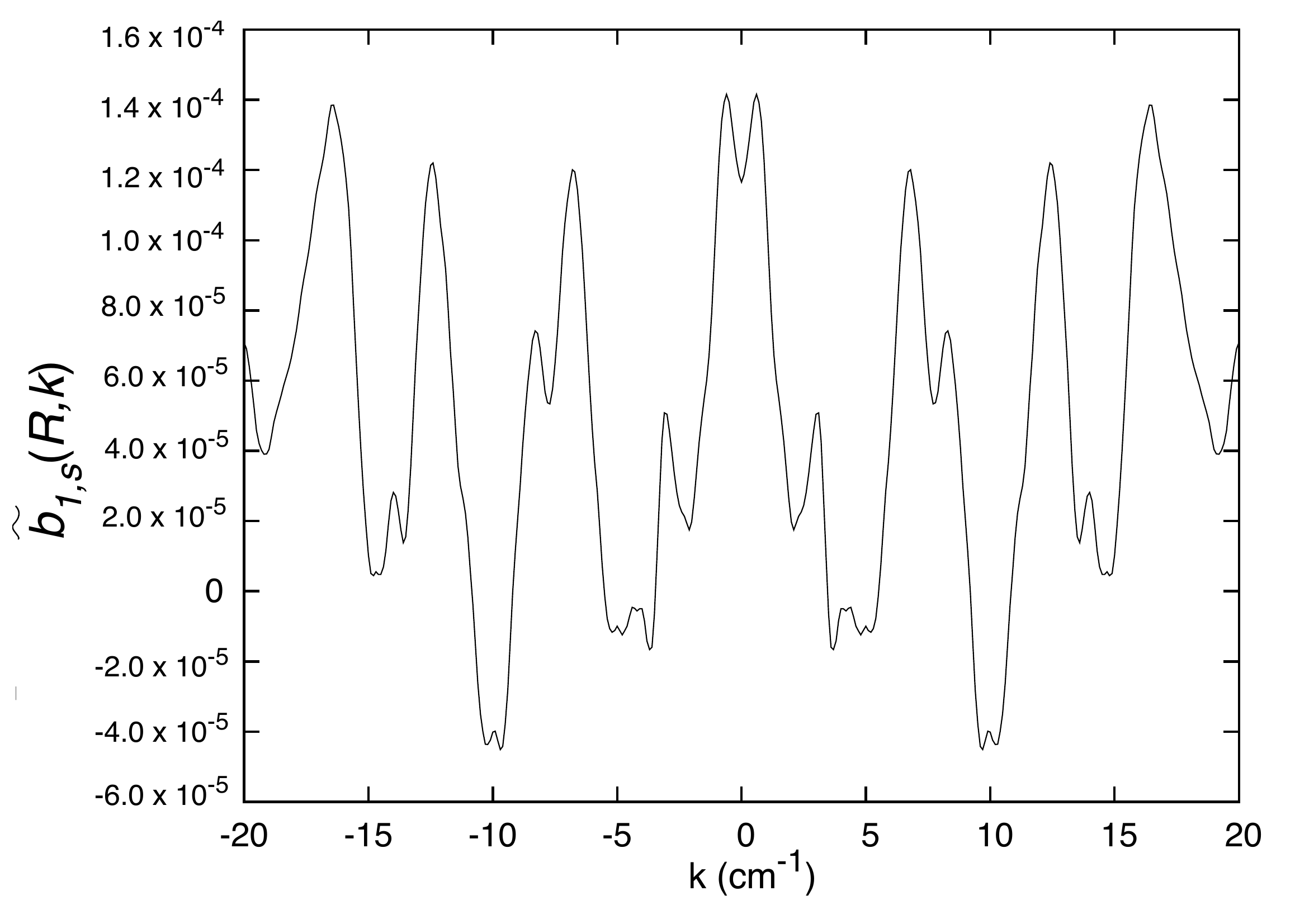}}
  \caption{\label{surfacenoise} The quantity $\tilde{b}_{1,s}(R,k)$ computed from uniform random noise (\ref{noisefield}) using a circular cylinder with $R=2$ cm.}
\end{figure}

\begin{figure}[hpt]
  \centering
  \resizebox{8.6cm}{!}{ \includegraphics*{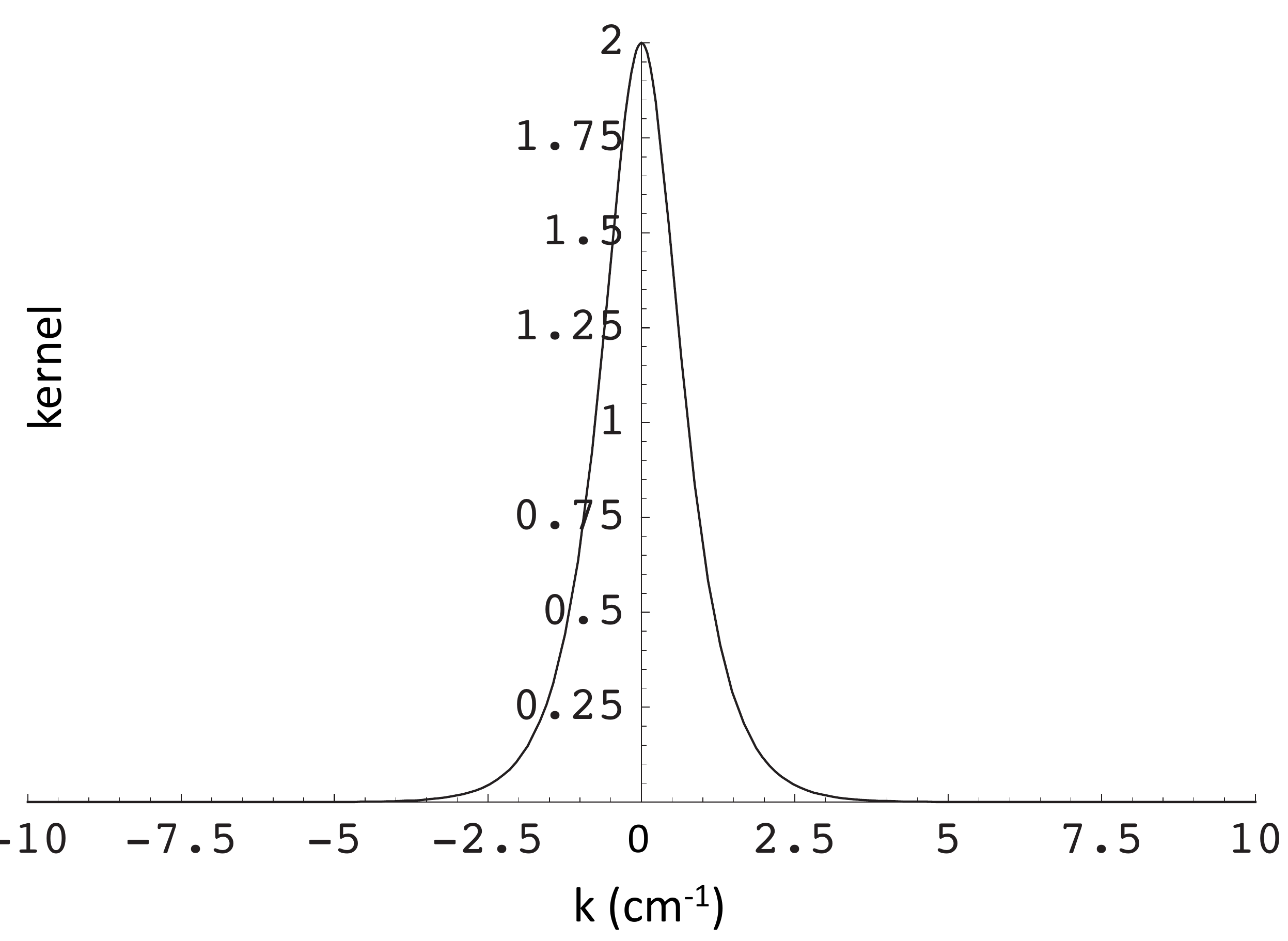}}
  \caption{\label{circkernel} The kernel $k^{m-1}/I_m^{\prime}(kR)$ as a function of $k$ for the case $m=1$ and $R=2$ cm. }
\end{figure}

\begin{figure}[hpt]
  \centering
  \resizebox{8.6cm}{!}{ \includegraphics*{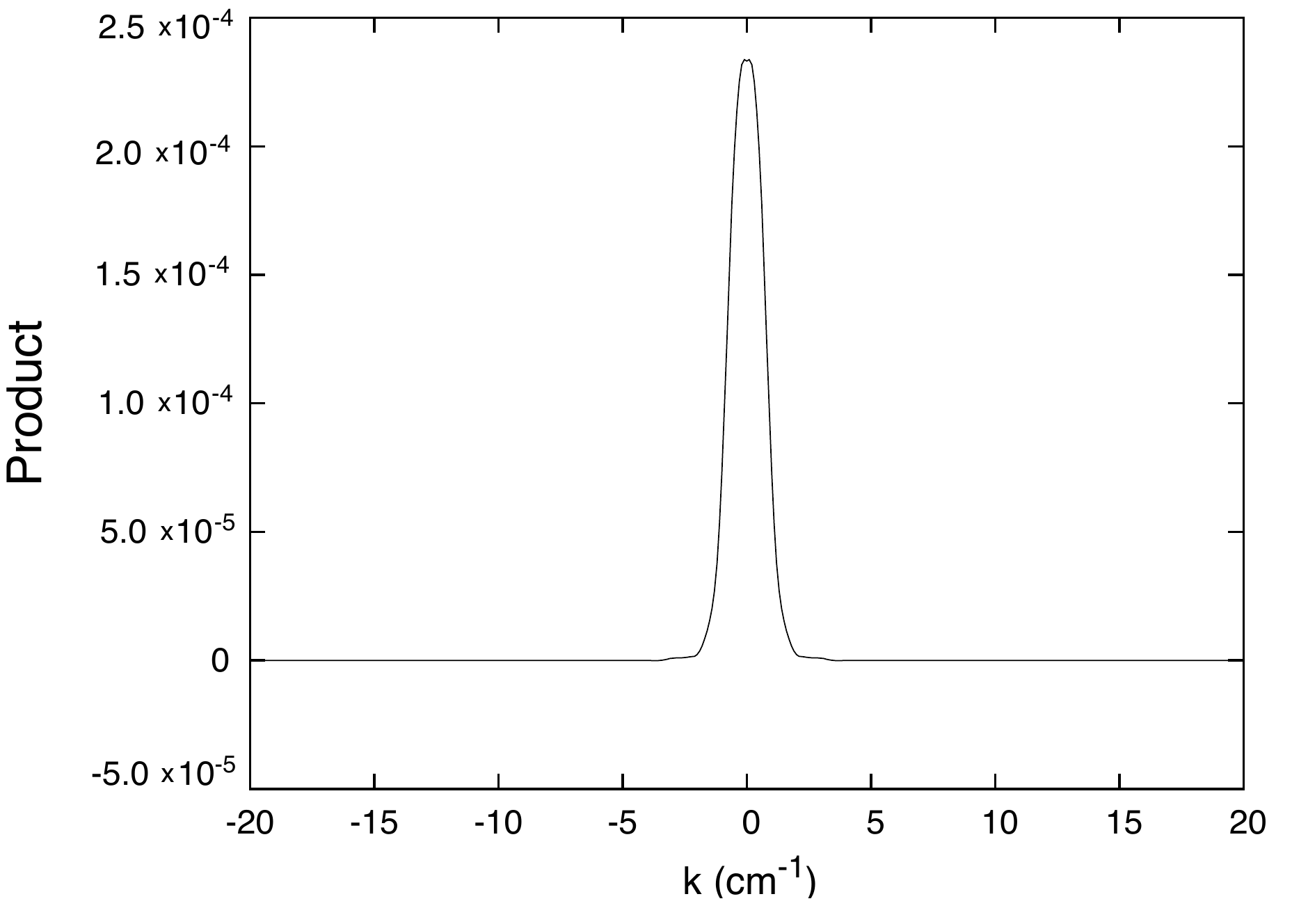}}
  \caption{\label{noiseproduct} The product $k^{m-1}\tilde{b}_{m,s}(R,k)/I_m^{\prime}$ appearing in (\ref{circulargradients}), as computed from uniform random noise (\ref{noisefield}) with $m=1$ and $R=2$  cm. }
\end{figure}

In Fig. \ref{surfacenoise}, we have displayed the computed quantity $\tilde{b}_{m,s}(R,k)$ appearing in (\ref{intermcyl}) for the case $m=1$.  It is a function of the spatial frequency $k$, having random variations of approximately uniform variance over the interval $[-20,20]  {\;}$cm$^{-1}$.  Fig. \ref{circkernel} displays the kernel $k^{m-1}/I^{\prime}_m(kR)$ multiplying $\tilde{b}_{m,s}(R,k)$ in (\ref{circulargradients}) for the case $m=1$ and $R=2$ cm.  Note the rapid decay of this function for large $|k|$.  Finally, Fig. \ref{noiseproduct} displays the product of these functions, illustrating the dramatic suppression of high-$k$ contributions to the Fourier integral appearing in (\ref{circulargradients}).  

\begin{figure}[hpt]
  \centering
  \resizebox{8.8cm}{!}{ \includegraphics*{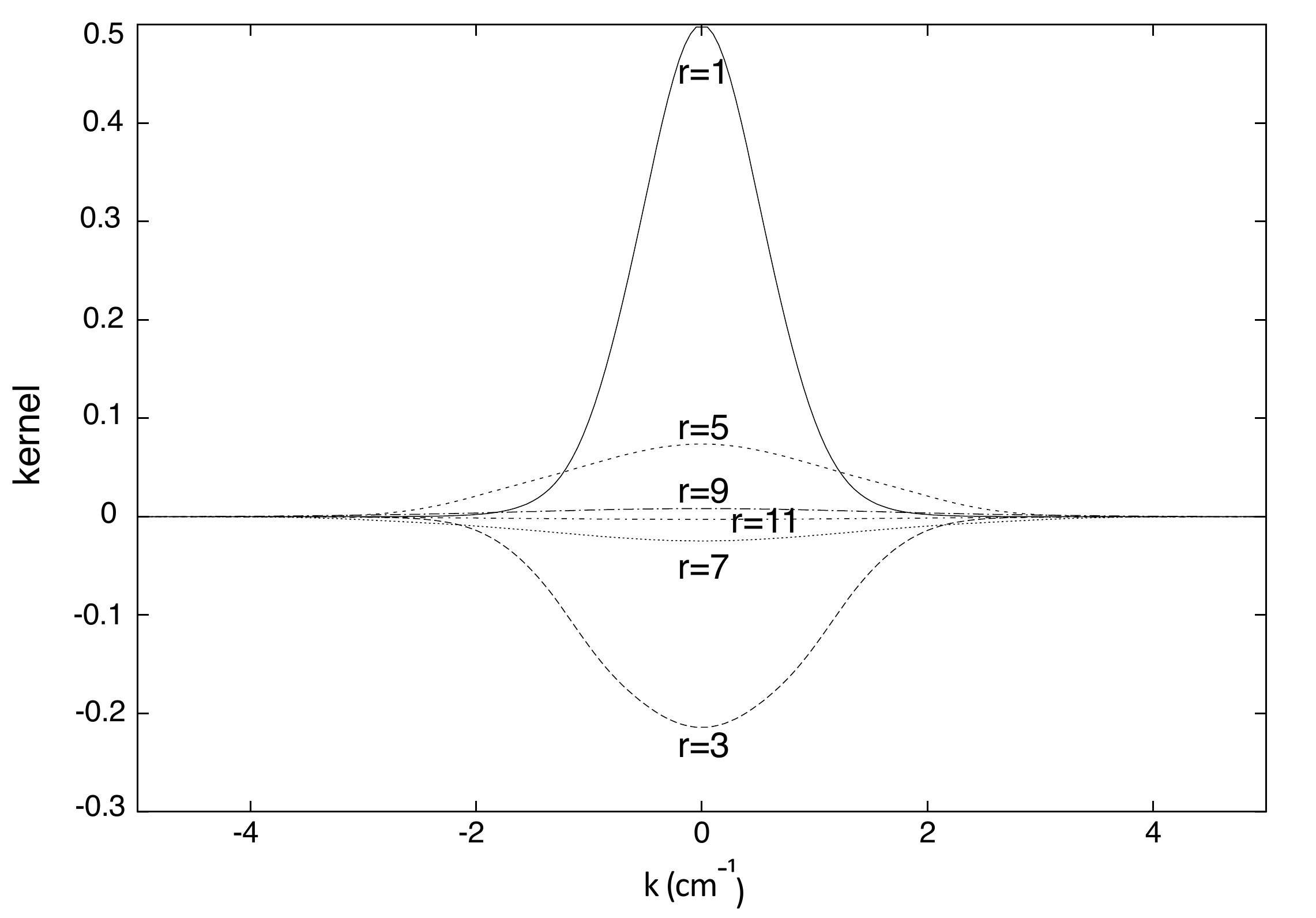}}
  \caption{\label{kernels1}The kernels $k^{m} \beta^r_m(k)/{\rm{Se}}_{r}^\prime(U,q)$ for the case $m=1$ and $r=1,3,5,7,9,11$, as a function of $k$, with $q$ and $k$ related by (\ref{qdef}).}
\end{figure}

A similar phenomenon occurs when fitting is performed using an elliptical cylinder.  In this case, a sequence of kernels contributes to (\ref{ellipticalgradients}) for each fixed $m$.  In Fig. \ref{kernels1} we have displayed the kernels contributing to the case $m=1$, with $x_{\rm max}=4$ cm and $y_{\rm max}=2$ cm.  Kernels take their maxima at $k=0$, and these maxima decrease monotonically with increasing index $r$.  All kernels decrease rapidly with increasing $|k|$ \cite{Chad, Dragt2}.

\begin{figure}
\resizebox{8.6cm}{!}{\includegraphics{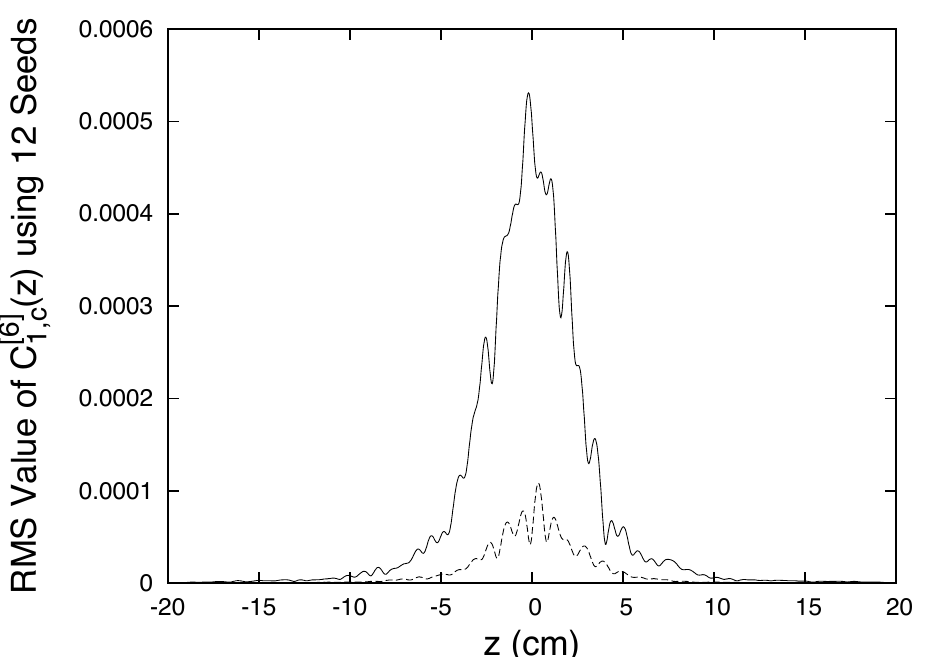}}
\caption{\label{Smooth1} The function $C_{1,c}^{[6]}$ as computed from a mesh containing random noise (\ref{noisefield}) at each mesh point.  Damping of this noise illustrates the effect of smoothing.  (Solid line) Result obtained using a circular cylinder with $R=2$ cm.  (Dashed line) Result obtained using an elliptical cylinder with semiminor axis of 2 cm and a semimajor axis of 4 cm.}
\end{figure}

\begin{figure}
\resizebox{8.6cm}{!}{\includegraphics{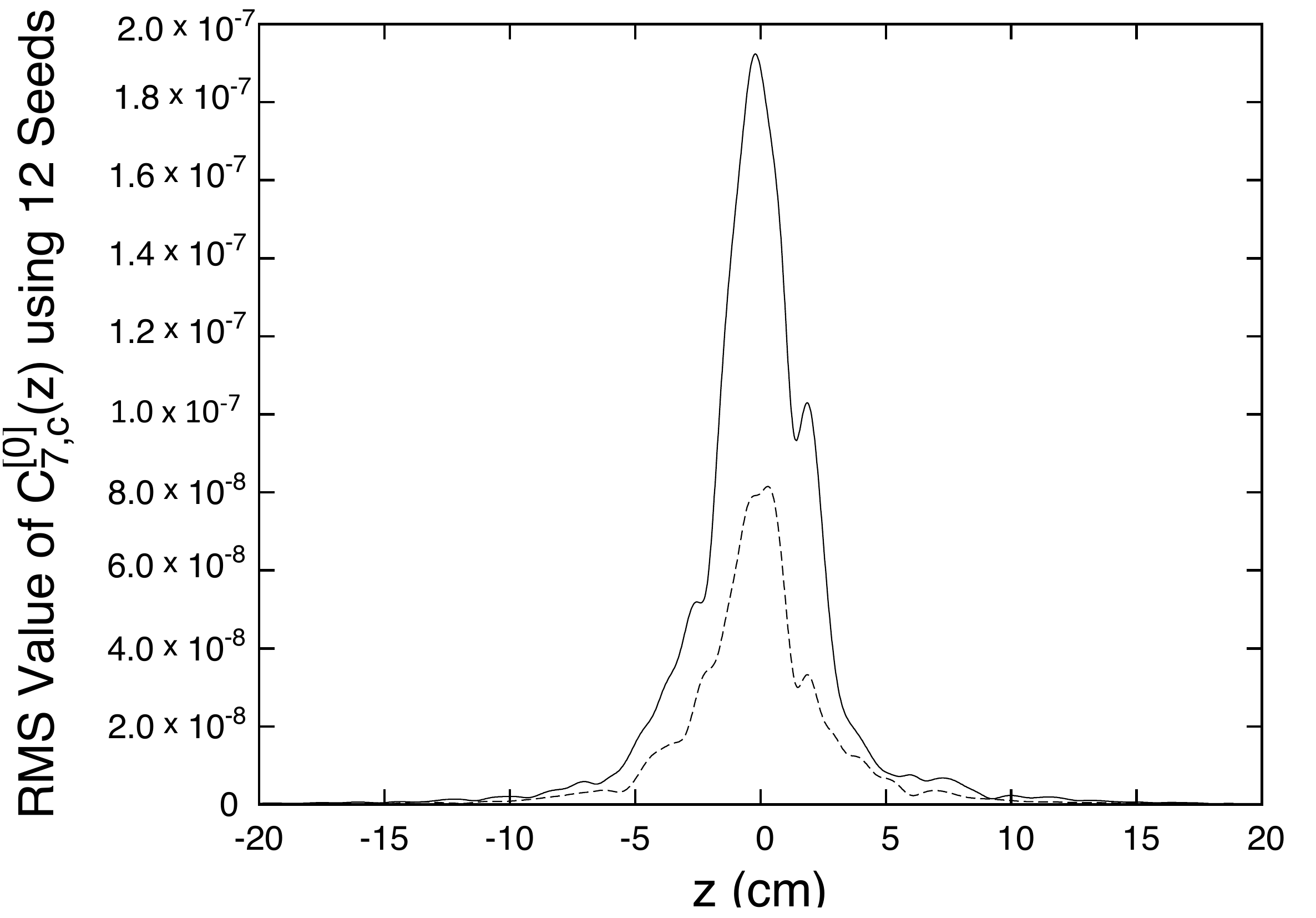}}
\caption{\label{Smooth2}  The function $C_{7,c}^{[0]}$ as computed from a mesh containing random noise (\ref{noisefield}) at each mesh point.  Damping of this noise illustrates the effect of smoothing.  (Solid line) Result obtained using a circular cylinder with $R=2$ cm.  (Dashed line) Result obtained using an elliptical cylinder with semiminor axis of 2 cm and a semimajor axis of 4 cm.}
\end{figure}

To study the effect of noise on the on-axis gradients, twelve distinct random fields were generated on a mesh according to (\ref{noisefield}).  Figs. \ref{Smooth1}-\ref{Smooth2} illustrate the on-axis gradients $C^{[6]}_{1,c}$ and $C^{[0]}_{7,c}$ as computed using these field values.  The solid line in Fig. \ref{Smooth1} illustrates the rms value of the on-axis gradient $C^{[6]}_{1,c}$, as computed using a circular cylinder of radius $R=2$ cm according to (\ref{circulargradients}).  The dashed line in Fig. \ref{Smooth1} illustrates the rms value of the on-axis gradient $C^{[6]}_{1,c}$, as computed using an elliptical cylinder of semiminor axis 2 cm and semimajor axis 4 cm according to (\ref{ellipticalgradients}).  In Fig. \ref{Smooth2}, similar results are shown for the on-axis gradient $C^{[0]}_{7,c}$.  

The attentive reader might wonder why we have displayed the $C^{[n]}_{m,c}$ for noise while the the monopole doublet field is governed by the gradients $C^{[n]}_{m,s}$.  The reason is that the $C^{[n]}_{m,s}$ are produced by fields that are predominantly in the vertical $(y)$ direction, and for such fields there is relatively less difference between circular and elliptical surface fitting because the semi-minor axis of the elliptical cylinder is the same as the radius of the circular cylinder.  Note also that in both cases only the component of the field normal to the surface is used.  Horizontal fields drive primarily the $C^{[n]}_{m,c}$.  However, in the  horizontal direction, the semi-major axis of the elliptical cylinder is substantially larger than the radius of the circular cylinder.   We therefore expect the advantage of using elliptical cylinders compared to circular cylinders will be most apparent in the  $C^{[n]}_{m,c}$.  Indeed, this is what Figs. 19 and 20 show.  The effects of errors in the surface data are suppressed more when the bounding surface is farther away from the field observation point.  As a result of this suppression, it is advantageous to use a fitting surface that is as far away as possible.

Suppose we compare the $C^{[n]}_{m,c}$ due to noise and shown in Figs. 19 and 20 with the corresponding $C^{[n]}_{m,s}$ in Figs. 12 and 13.  This is reasonable because a horizontal monopole doublet would produce $C^{[n]}_{m,c}$ analogous to the $C^{[n]}_{m,s}$ shown in Figs. 12 and 13.  Remarkably, we find that, as a result of smoothing, the effect of 1\% noise in the field data produces, on average, only on the order of 0.01\% changes in the on-axis gradients.  


\section{\label{sec:appl}Applications}
\subsection{\label{sec:appl:DR} ILC Damping Ring Wiggler Fields}
A less stringent test of the accuracy of surface methods is that the magnetic field, as computed from surface data using field values on a 3-dimensional mesh, should reproduce the magnetic field at the interior mesh points.  (This is also a test of the quality of the magnetic data on the mesh.)  We computed such an interior fit, and the associated transfer map, for the modified CESR-c design of the Cornell wiggler, which has been adopted as the design prototype for use in International Linear Collider studies \cite{ILC, Urban}.  Cornell provided data obtained from the 3-dimensional finite element modeling code OPERA-3d for the field components $B_x$, $B_y$, and $B_z$ on a mesh of spacing $0.4\times0.2\times0.2$ cm in a volume $10.4\times5.2\times480$ cm, extending beyond the fringe-field region.  The field components are provided to a precision of 0.1 G relative to a peak field of 16.7 kG.  An elliptic cylinder with semimajor axis $4.4$ cm and semiminor axis $2.4$ cm was placed in the domain of the data, and the field on the elliptic cylinder boundary was constructed using bicubic interpolation.  See Figs. \ref{ellipticalcylindergraphic} and \ref{mesh}.

\begin{figure}
\begin{center}
\resizebox{8.6cm}{!}{\includegraphics{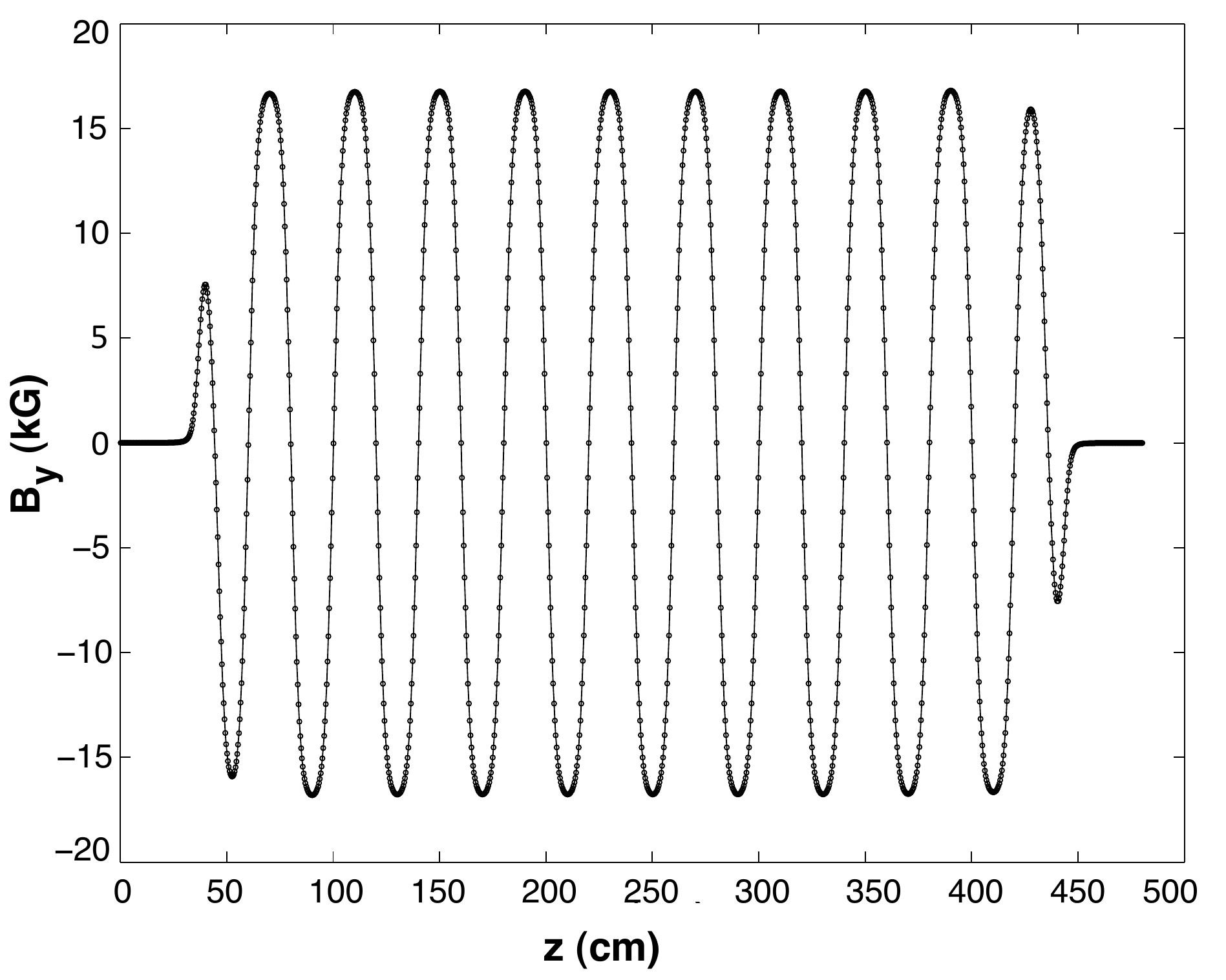}}
\end{center}
\caption{\label{wigglerByinz} Fit to the proposed ILC wiggler vertical field versus $z$, where $x=0.4$ cm, $y=0.2$ cm.  The solid line is computed using data on the surface of an elliptical cylinder with  $x_{max}=4.4$ cm, $y_{max}=2.4$ cm, using the polynomial series for $\mbox{\boldmath $B$}$ obtained from (\ref{psieq}) or (\ref{Atran}).  Dots represent numerical data provided by OPERA-3d.}
\end{figure}

\begin{figure}
\begin{center}
\resizebox{9.0cm}{!}{\includegraphics{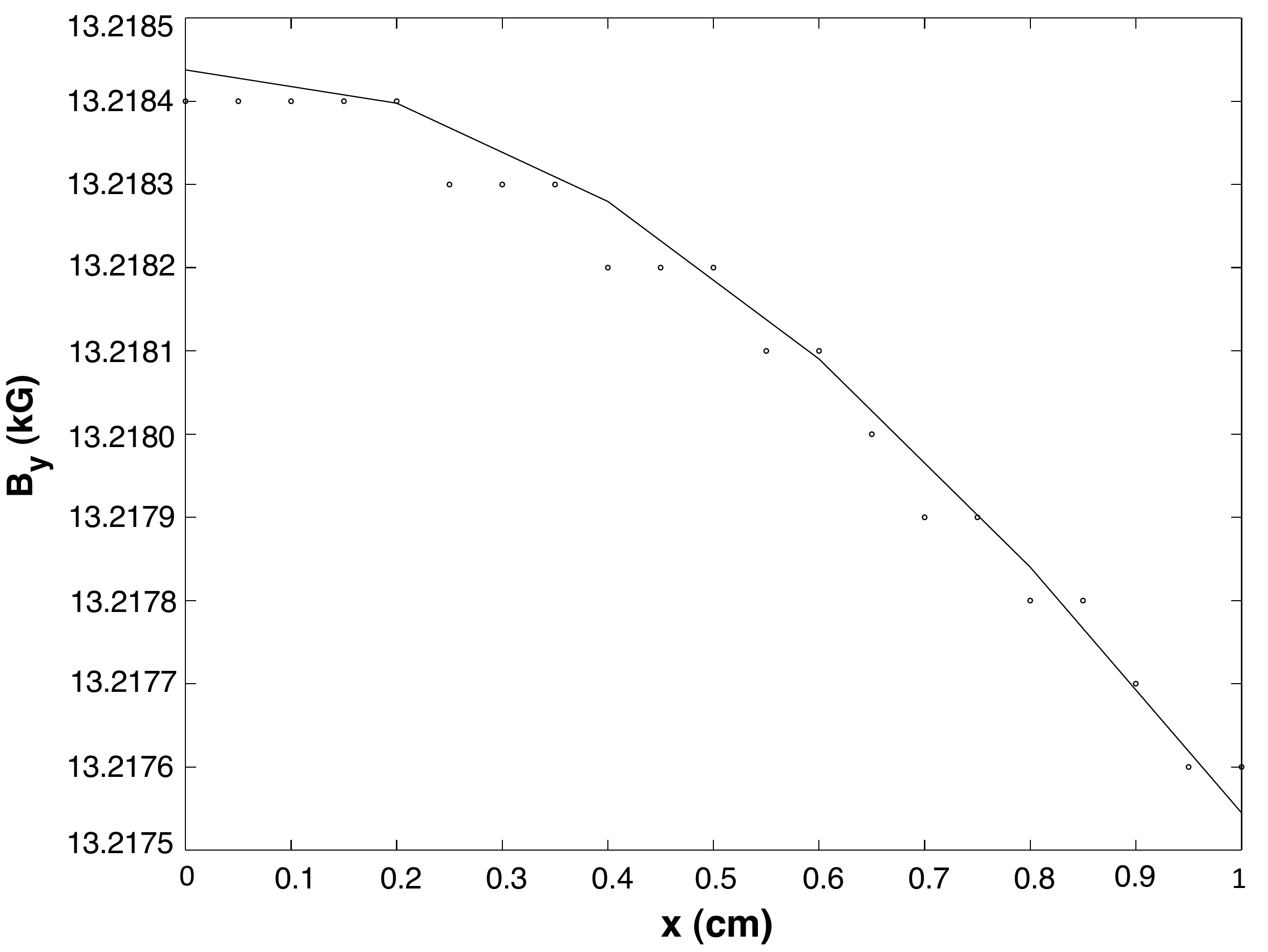}}
\end{center}
\caption{\label{wigglerByinx} Fit to the proposed ILC wiggler vertical field versus $x$, where $y=0.1$ cm, $z=104.2$ cm.  The solid line is computed using data on the surface of an elliptical cylinder with  $x_{max}=4.4$ cm, $y_{max}=2.4$ cm, using the polynomial series for $\mbox{\boldmath $B$}$ obtained from (\ref{psieq}) or (\ref{Atran}).  Dots represent numerical data provided by OPERA-3d.}
\end{figure}

\begin{figure}
\begin{center}
\resizebox{8.5cm}{!}{\includegraphics[height=2in,width=2.36in]{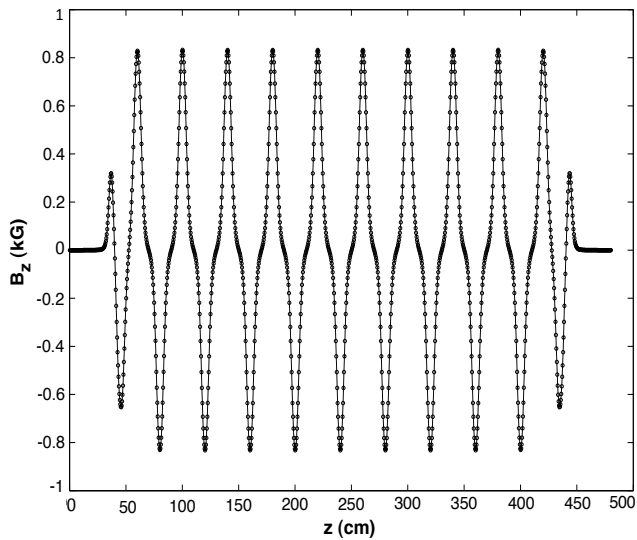}}
\end{center}
\caption{\label{wigglerBzinz} Fit to the proposed ILC wiggler longitudinal field versus $z$, where $x=0.4$ cm, $y=0.2$ cm.  The solid line is computed using data on the surface of an elliptical cylinder with  $x_{max}=4.4$ cm, $y_{max}=2.4$ cm, using the polynomial series for $\mbox{\boldmath $B$}$ obtained from (\ref{psieq}) or (\ref{Atran}).  Dots represent numerical data provided by OPERA-3d.}
\end{figure}

The interior field was computed using the on-axis gradients through terms of degree 6 in $x,y$ over the domain of the original data.  This solution for the interior field was then compared to the original data at each grid point.  Fig. \ref{wigglerByinz} displays the vertical field $B_y$ off-axis at $(x,y)=(0.4,0.2)$ cm along the length of the wiggler.
The field data (points) are shown along with computed values (solid line).  Note that the fitted field captures the fringe-field behavior.  The relative error was found to satisfy the bound $\delta{|\mbox{\boldmath $B$}_{data}-\mbox{\boldmath $B$}_{fit}|}/|\mbox{\boldmath $B$}|_{peak}\le3.5\times{10^{-4}}$.  We observe that this error is comparable to that found for the monopole-doublet benchmark.  Presumably it is due to errors in numerical integration, errors in interpolating onto the elliptic cylinder, errors arising from neglecting terms beyond degree 6, etc., as well as possible failure of the OPERA-3d data to be Maxwellian.
Fig. \ref{wigglerByinx} illustrates the horizontal roll-off of the vertical field at $y=0.1$ cm, $z=104.2$ cm for $0\leq{x}\leq{1}$ cm.  Note the discrete jumps in the original data, reflecting the number of digits retained in the output of the numerical computation.  Despite the small variation of $B_y$ in $x$, the fit goes through the interior data.  Finally, Fig. \ref{wigglerBzinz} illustrates the longitudinal field $B_z$, again at $(x,y)=(0.4,0.2)$ cm along the wiggler.  Note that no information about $B_z$ was used to generate this field, since only the component of $\mbox{\boldmath $B$}$ normal to the elliptic cylinder surface was used to generate the interior solution.  

\begin{figure}
\begin{center}
\resizebox{9.6cm}{!}{\includegraphics{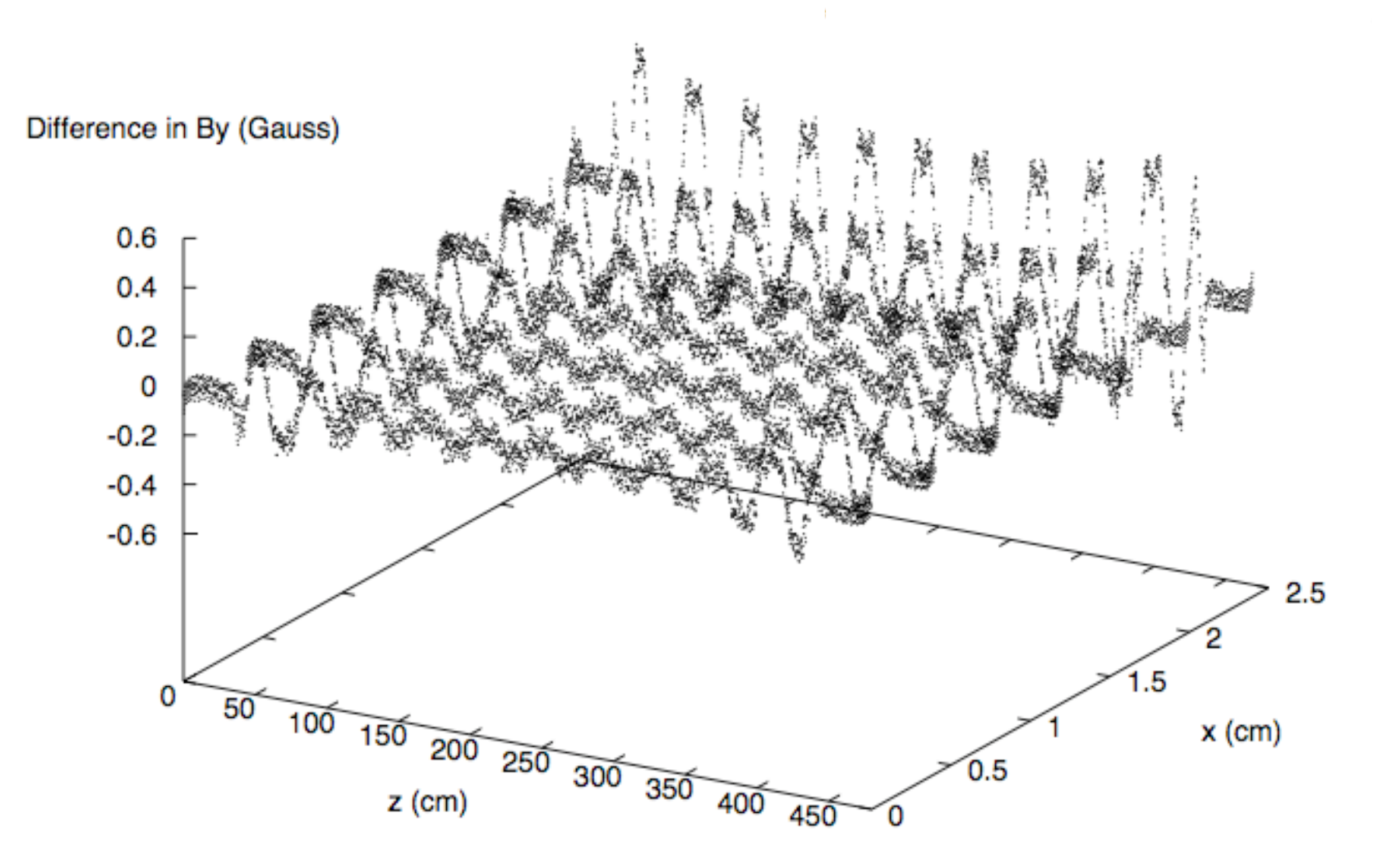}}
\caption{\label{residuals}  Difference (Gauss) between the vertical field $B_y$ of the proposed ILC wiggler and its fitted value across the midplane $y=0$.  Peak field is 16.7 kG. }
\end{center}
\end{figure}

The error for $B_y$ on-axis lies in the range $0.1$-$0.2$ G along the length of the wiggler, increasing slightly near the end poles.  A plot of residuals in the plane $y=0$ is displayed in Fig. \ref{residuals}.  Note that the error is within $0.6$ G over this region of the $x$-$z$ plane.  The error begins to increase rapidly at about $x=2$ cm; this may be due to retaining only terms through degree 6 in the on-axis gradients, or perhaps a finite domain of convergence of the associated power series for $B_y(x,y,z)$.  For all $|x|\leq 2$ cm, the peak error is $0.3$ G.  We remark that this peak error amounts to a relative error of less than 2 parts in $10^5$, which is remarkably small compared to the error for the data of Fig. 21.  We expect the error to behave like a harmonic function, and therefore it should grow as one approaches the boundary.  Conversely, it should be the smallest on the center line.  The observed error follows this pattern.  Finally, this phenomenon may also be a factor in the observed increase in the error at and beyond
$x=2$ cm.

\subsection{\label{sec:appl:DR} ILC Damping Ring Wiggler Design Orbit and Associated Transfer Map}

The on-axis gradients computed for the ILC wiggler were then used in the code MaryLie/IMPACT to integrate, simultaneously, i) equations for the design orbit of a 5 GeV positron through the wiggler, ii) equations for the matrix elements of the linear part of the transfer map through the wiggler, and iii) equations for the coefficients of the generating polynomials $f_3$,$f_4$,... appearing in the Lie factorization of the transfer map.  Each generator of the symplectic transfer map ${\mathcal M}$ is computed in variables representing deviations from the design orbit.  See the Appendix for a brief discussion of Lie methods.

\begin{figure}
\begin{center}
\resizebox{8.6cm}{!}{\includegraphics[height=3.09in,width=4.0in]{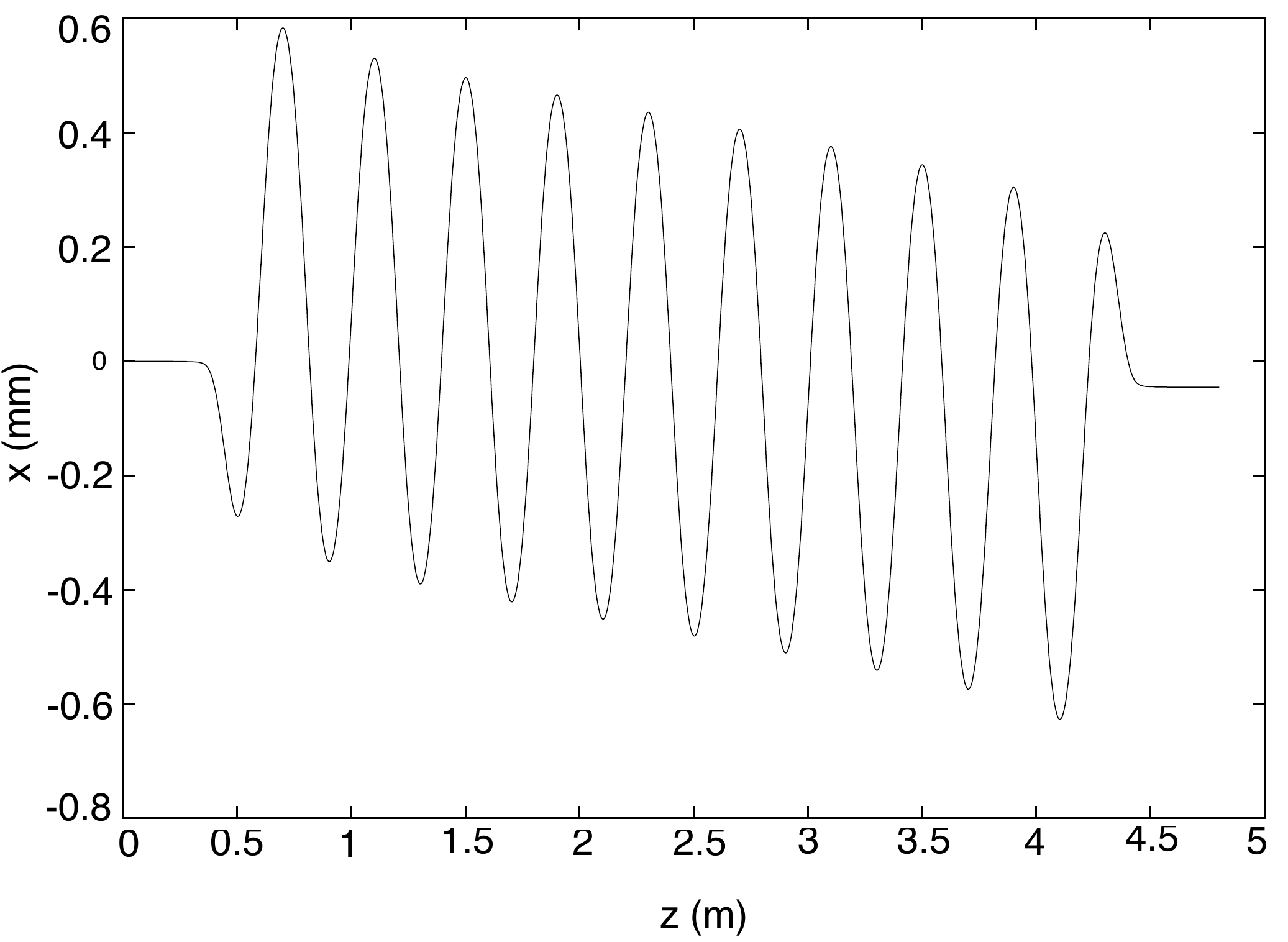}}
\resizebox{8.6cm}{!}{\includegraphics[height=3.09in,width=4.0in]{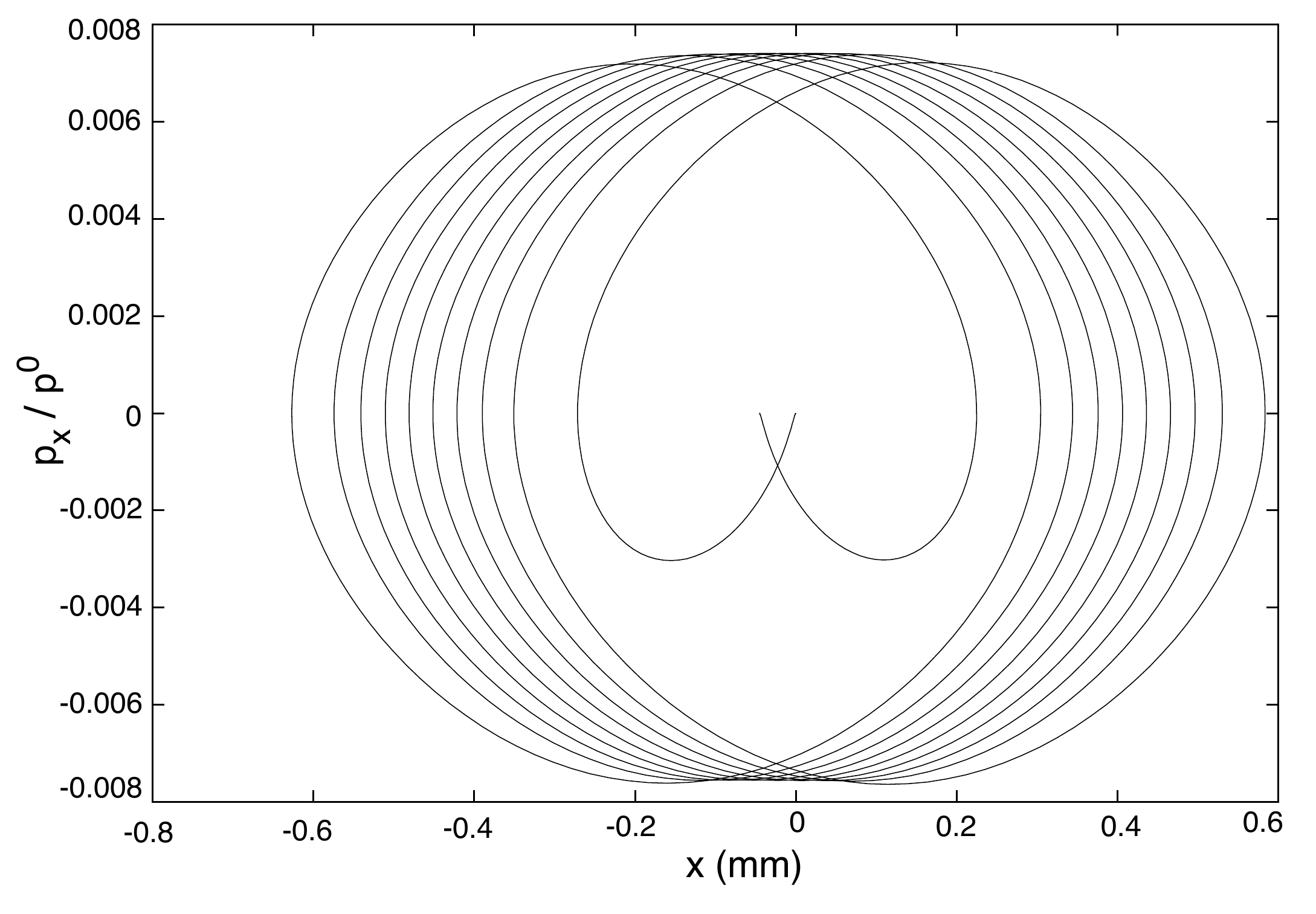}}
\caption{\label{reftrajectory} Design orbit for a $5$ GeV positron through the proposed ILC wiggler.  (Upper) Coordinate $x$(mm) along the length of the wiggler $z$(m).  (Lower) Design orbit in the phase space defined by coordinates $x$(mm) and $p_x/p^0$.}
\end{center}
\end{figure}

The design orbit is displayed in Fig. \ref{reftrajectory}, and Table I lists its initial and final conditions.
Table II displays the matrix $R_2$ that describes the linear part of the transfer map in (\ref{factorization}).  Phase-space coordinates are arranged in the order $(x,p_x,y,p_y,\tau,p_\tau)$.  The first few Lie generators $f_m$ of the nonlinear part of the transfer map are listed in
Table III.  

\begin{table*}[htbp]
\begin{center}
\small
\caption{Design Orbit Specifications.}
\begin{tabular}{ccc}
\hline
$E$ [GeV] & $p^0$ [GeV/c] & $B\rho$ [Tm]\\
$5.00000000000000$ & $5.000510972950664$ & $16.67990918220719$ \\
\hline
& Entry ($z=0$ m) & Exit ($z=4.8$ m) \\
\hline
$x$ [cm] & 0.0  & -0.00449920546655753\\
$p_x^{mech}/p^0$ & $ 0.0$  & 4.5494526506306504$\times 10^{-10}$\\
$y$ [cm] &  0.0  &  0.0\\
$p_y^{mech}/p^0$ &  0.0 &  0.0\\
$ct$ [cm] &  0.0  & 480.00494875303383\\
$p_t/p^0c$ & $-1.0000000052213336$  & -1.0000000052213336\\
\hline
\end{tabular}\\
\end{center}
\label{table:parameters}
\end{table*}

\begin{table*}[htbp]
\begin{center}
\small
\caption{Linear transfer map $R_2$ for the ILC damping ring wiggler.}
\begin{tabular}{cccccc}
$1.000056$ & $4.800233\times 10^2$ & $0.000000$ & $0.000000$ & $0.000000$ & $-4.500058\times 10^{-3}$ \\[1 ex]
$2.235501\times 10^{-7}$ & $1.000052 $ & $0.000000$ & $0.000000$ & $0.000000$ & $-1.001075\times 10^{-9}$ \\[1 ex]
$0.000000$ & $0.000000$ & $9.404373\times 10^{-1}$ & $4.687866\times 10^2$ & $0.000000$ & $0.000000$ \\[1ex]
$0.000000$ & $0.000000$ & $-2.465383\times 10^{-4}$ & $9.404414\times 10^{-1}$ & $0.000000$ & $0.000000$ \\[1 ex]
$-4.857478\times10^{-14}$ & $-4.499810\times 10^{-3}$ & $0.000000$ & $0.000000$ & $1.000000$ & $9.897806\times 10^{-3}$\\[1ex]
$0.000000$ & $0.000000$ & $0.000000$ & $0.000000$ & $0.000000$ & $1.000000$\\[1ex]
\end{tabular}
\end{center}
\label{table:mapmatrix}
\end{table*}

\begin{table}[h]
\caption{First few nonvanishing Lie generators $f_m$ for the ILC damping ring wiggler.}
\centering
\begin{tabular}{ccc}
\hline
Index  & Monomial & Coefficient  \\[0.5ex]
\hline\\
  28 & $x^3$ &-1.0738513995490168$\times 10^{-9}$\\[1ex]
  29 & $x^2 p_x$ & 4.630793805976143$\times 10^{-4}$\\[1ex]
  33 & $x^2 p_{\tau}$ & 1.1070545309022173$\times 10^{-7}$\\[1ex]
  34 & $x p_x^2$ & -2.2214155912078926$\times 10^{-1}$ \\[1ex]
   $\cdots$ & $\cdots$ & $\cdots$ \\[1ex]
  209 & $p_{\tau}^4$ & -4.949353522256519$\times 10^{-3}$\\[1ex]
\hline
\end{tabular}
\label{table:generators}
\end{table}

\section{\label{conclude} Concluding Summary}

Surface methods provide a reliable and numerically robust technique for extracting transfer maps from numerical field data.  By benchmarking them against a numerically challenging  problem whose results are known exactly, we have verified that surface methods have all the advantages claimed in the beginning of Section III.  In particular, we demonstrated that errors of a few parts in $10^4$ can be achieved for all on-axis gradients $C^{[n]}_{m,\alpha}(z)$ required to compute transfer maps through $7^{\rm{th}}$ order, and that the results obtained were remarkably insensitive to noise.  Moreover, these small errors can be further reduced if desired with the aid of a finer grid. 

Subsequently we applied surface methods to compute the interior field for the proposed ILC Damping Ring wigglers.  Consistent with the accuracy displayed by the monopole-doublet benchmark results, excellent fits were demonstrated for interior fields.  We also illustrated the computation of the design orbit and its associated  transfer map based on surface methods.  

In summary, the use of surface methods makes it possible, for the first time, to compute for straight beam-line elements realistic transfer maps for real magnets including all fringe and high-order multipole field effects.

\begin{figure}[hpt]
  \centering
    \resizebox{8.6cm}{!}{ \includegraphics*{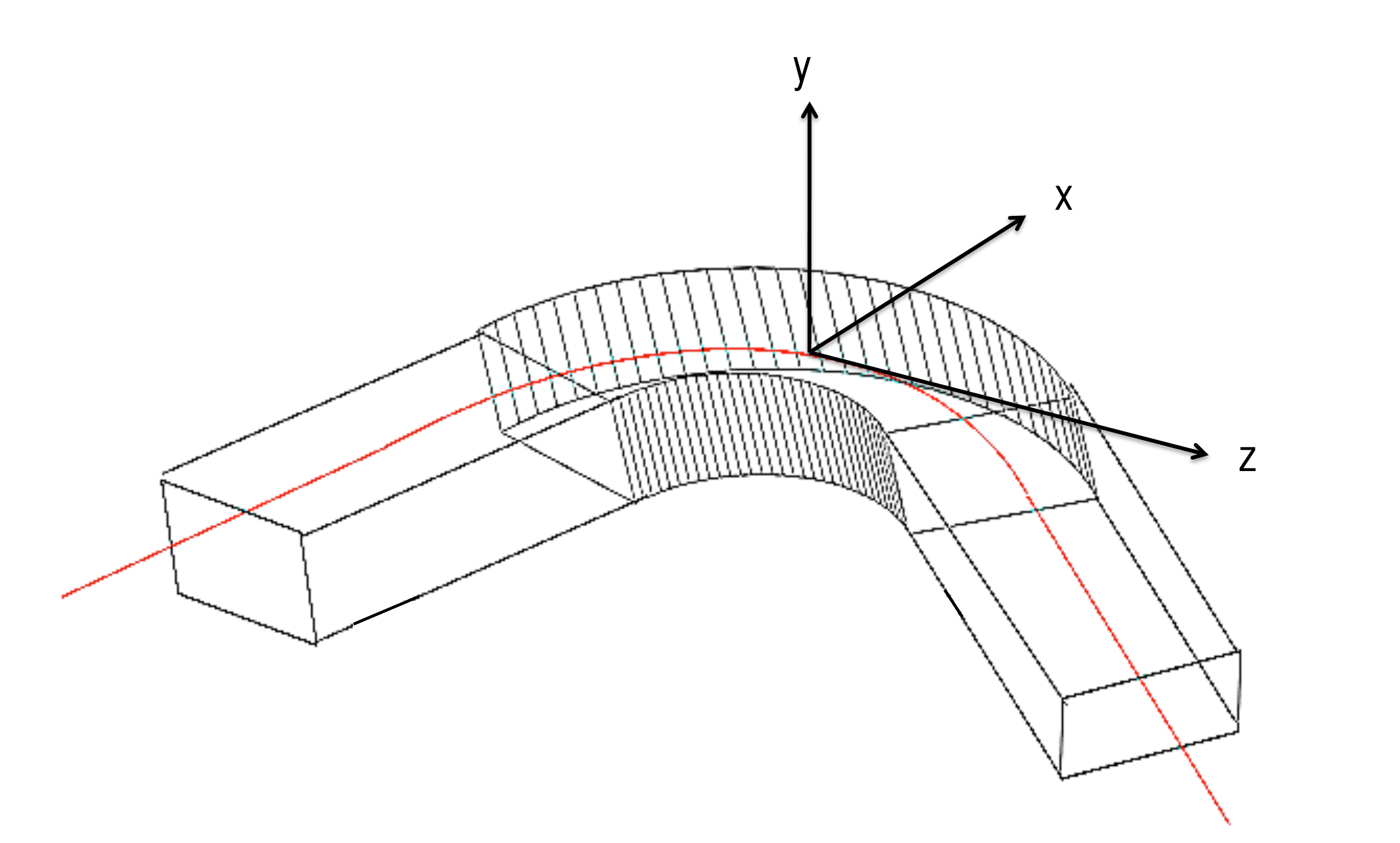}}
  \caption{\label{bentbox} (Color) Illustration of a bent box with straight ends, used for computing a transfer map for bending dipoles with large sagitta.  A design orbit is illustrated in red.}
\end{figure}

In many cases, however, we are interested in magnetic elements with significant sagitta, such as dipoles with large bending angles.  In these cases, it is not possible in general to surround the design orbit with a cylindrical surface that lies interior to all iron or other magnetic sources.  Part II of this paper will describe an alternative, but more computationally intensive, method suitable for general geometries including that of a bent box with straight ends.  See Fig. \ref{bentbox}.  In this case we obtain simple, geometry-independent kernels for computing the interior vector potential and its derivatives.  All the advantages demonstrated in this paper for surface methods will be retained.  

\appendix

\section*{APPENDIX}
MaryLie/IMPACT is a hybrid code that utilizes Lie-algebraic methods for computing and manipulating charged-particle transfer maps through $5^{\rm{th}}$ order, while space-charge effects are treated using Particle-in-Cell methods \cite{MLI}.  We made use of its Lie algebraic capabilities.  In the Lie algebraic approach, maps are computed and manipulated in Lie algebraic form.  Each map describes the transformation of the full six-dimensional phase-space coordinates of a particle as it passes through a given beam-line element.  Because of the symplectic nature of Hamiltonian motion, through aberrations of order $(n-1)$ such a map has the Lie representation 
\begin{equation}
{\cal M} = {\cal R}_2 \exp (:f_3:) \exp (:f_4:) \cdots \exp (:f_n:) \label{factorization}
\end{equation}
where ${\cal R}_2$ describes the linear part of the map, and each $f_j$ is a homogeneous polynomial of degree $j$ describing the nonlinear part of the map \cite{Dragt2, Dragt4}.

The linear map ${\cal R}_2$ and the Lie generators
$f_j$ are determined by solving the equation of motion
\begin{equation}
\dot{{\cal M}} = {\cal M}:-H:  \label{mapequation}
\end{equation}
where
\begin{equation}
H = H_2 + H_3 + H_4 + \cdots  \label{Hseries}
\end{equation}
is the charged-particle Hamiltonian expressed in terms of deviation variables about the design orbit and expanded in a homogeneous polynomial series.  The deviation variable Hamiltonian $H$ is
determined in turn by the Hamiltonian $K$ for which path length is the
independent variable.  In Cartesian coordinates and with $z$ taken as the
independent variable, $K$ is given by the relation
\begin{align}
K =& -[(p_t + q\Phi )^2/c^2 - m^2c^2 - (p_x - qA_x)^2 \nonumber \\
&- (p_y - qA_y)^2]^{1/2}-qA_z.  \label{Keq}
\end{align}
Here $\Phi$ and $\mbox{\boldmath $A$}$ are the electric scalar and magnetic vector potentials, respectively. 

We conclude that (in the case of no electric fields, $\Phi = 0$) what is needed are Taylor expansions for the components of $\mbox{\boldmath $A$}$ in the deviation variables $x$ and $y$.  In the straight-element case these Taylor expansions can be found using the circular cylinder harmonic based expansion (\ref{Atran}).
Suppose, for example, we wish to retain in the expansion of the Hamiltonian $H$ appearing in (\ref{Hseries}) homogeneous polynomials through degree $8$.  That is what is required to compute transfer maps through $7^{\rm{th}}$ order.  If the design orbit lies on the $z$ axis, as will be the case for any straight multipole such as a solenoid, quadrupole, sextupole, octupole, etc., this expansion is straightforward because in this case the Cartesian coordinates $x,y$ are already deviation variables.  We see from (\ref{Keq}) that we must retain homogeneous polynomials in the variables $x,y$ through degree $7$ in the expansions of $A_x$ and $A_y$, and homogeneous polynomials in the variables $x,y$ through degree $8$ in the expansion of $A_z$.  Inspection of (\ref{Atran}) shows that for the cases $m=0$ or $m$ odd we then need the $C^{[n]}_{m,\alpha}(z)$ with $(m+n)\le7$.  For the cases of even $m$ we need the $C^{[n]}_{m,\alpha}(z)$ with $(m+n)\le8$. 

In the case of a wiggler, the design orbit oscillates around the $z$ axis.  See Fig. 25. Now the requirement that the Hamiltonian $H$ appearing in (\ref{Hseries}) be an expansion in deviation variables about the design orbit is more involved: The components of $\mbox{\boldmath $A$}$ must be expanded about the design orbit.  As they stand in (\ref{Atran}), they are expanded about the $z$ axis.  If we are to retain terms in $H$ through degree 8 when a re-expansion is made about the design orbit, then coefficients $C^{[n]}_{m,\alpha}$ beyond those listed in the previous paragraph must in principle be included in the calculation.  We say that higher-degree terms produce lower-degree terms due to {\em feed down}.  However, the effect of feed down is small if the oscillations are of small amplitude, as they are for the proposed ILC wiggler.  In this example we have found that the approximation of retaining only the terms in $A_x$, $A_y$, and $A_z$ through degree 8 produces changes in the design orbit and the transfer map about the design orbit that are comparable to or smaller than the errors found in the benchmark studies of Section IV.  We also remark that this feed-down problem does not arise when the general geometry methods to be described in Part II are employed.

\bibliography{Final_BV_StraightRev}

\begin{thebibliography}{12}
\expandafter\ifx\csname natexlab\endcsname\relax\def\natexlab#1{#1}\fi
\expandafter\ifx\csname bibnamefont\endcsname\relax
  \def\bibnamefont#1{#1}\fi
\expandafter\ifx\csname bibfnamefont\endcsname\relax
  \def\bibfnamefont#1{#1}\fi
\expandafter\ifx\csname citenamefont\endcsname\relax
  \def\citenamefont#1{#1}\fi
\expandafter\ifx\csname url\endcsname\relax
  \def\url#1{\texttt{#1}}\fi
\expandafter\ifx\csname urlprefix\endcsname\relax\def\urlprefix{URL }\fi
\providecommand{\bibinfo}[2]{#2}
\providecommand{\eprint}[2][]{\url{#2}}

\bibitem[{\citenamefont{Hildebrand}(1987)}]{Hildebrand}
\bibinfo{author}{\bibfnamefont{F.~B.} \bibnamefont{Hildebrand}},
  \emph{\bibinfo{title}{Introduction to Numerical Analysis}}
  (\bibinfo{publisher}{Dover Books}, \bibinfo{year}{1987}),
  \bibinfo{edition}{2nd} ed., \bibinfo{note}{section 3.3, p. 85. He writes ``In
  particular, numerical differentiation should be avoided wherever possible
  $\cdots$''}.

\bibitem[{\citenamefont{Abell}(2006)}]{Abell}
\bibinfo{author}{\bibfnamefont{D.}~\bibnamefont{Abell}},
  \bibinfo{journal}{Phys. Rev. ST Accel. Beams} \textbf{\bibinfo{volume}{9}},
  \bibinfo{pages}{052001} (\bibinfo{year}{2006}).

\bibitem[{\citenamefont{Mitchell}(2007)}]{Chad}
\bibinfo{author}{\bibfnamefont{C.~E.} \bibnamefont{Mitchell}}, Ph.D. thesis,
  \bibinfo{school}{University of Maryland, College Park}
  (\bibinfo{year}{2007}),
  \bibinfo{note}{\texttt{http://www.physics.umd.edu/dsat}}.

\bibitem[{\citenamefont{Dragt}(2009)}]{Dragt2}
\bibinfo{author}{\bibfnamefont{A.~J.} \bibnamefont{Dragt}},
  \emph{\bibinfo{title}{Lie Methods for Nonlinear Dynamics with Applications to
  Accelerator Physics}} (\bibinfo{publisher}{University of Maryland},
  \bibinfo{year}{2009}),
  \bibinfo{note}{\texttt{http://www.physics.umd.edu/dsat}}.

\bibitem[{\citenamefont{Venturini and Dragt}(1999)}]{Venturini}
\bibinfo{author}{\bibfnamefont{M.}~\bibnamefont{Venturini}} \bibnamefont{and}
  \bibinfo{author}{\bibfnamefont{A.}~\bibnamefont{Dragt}},
  \bibinfo{journal}{Nucl. Inst. and Meth. A} \textbf{\bibinfo{volume}{427}},
  \bibinfo{pages}{387} (\bibinfo{year}{1999}).

\bibitem[{\citenamefont{Morse and Feshbach}(1953)}]{Feshbach}
\bibinfo{author}{\bibfnamefont{P.}~\bibnamefont{Morse}} \bibnamefont{and}
  \bibinfo{author}{\bibfnamefont{H.}~\bibnamefont{Feshbach}},
  \emph{\bibinfo{title}{Methods of Theoretical Physics}}, vol.
  \bibinfo{volume}{1 \& 2} (\bibinfo{publisher}{McGraw-Hill Book Company,
  Inc.}, \bibinfo{year}{1953}).

\bibitem[{\citenamefont{Weisstein}()}]{Ellfigure}
\bibinfo{author}{\bibfnamefont{E.}~\bibnamefont{Weisstein}},
  \emph{\bibinfo{title}{{\it Mathworld} ${\;}$-${\;}$ a {Wolfram} web
  resource}}, \bibinfo{note}{\texttt{http://mathworld.wolfram.com/ \\
  EllipticCylindricalCoordinates.html}}.

\bibitem[{\citenamefont{McLachlan}(1964)}]{McLachlan}
\bibinfo{author}{\bibfnamefont{N.~W.} \bibnamefont{McLachlan}},
  \emph{\bibinfo{title}{Theory and Application of Mathieu Functions}}
  (\bibinfo{publisher}{Dover Publications, Inc.}, \bibinfo{year}{1964}).

\bibitem[{\citenamefont{Phinney et~al.}(2007)\citenamefont{Phinney, Toge, and
  Walker}}]{ILC}
\bibinfo{editor}{\bibfnamefont{N.}~\bibnamefont{Phinney}},
  \bibinfo{editor}{\bibfnamefont{N.}~\bibnamefont{Toge}}, \bibnamefont{and}
  \bibinfo{editor}{\bibfnamefont{N.}~\bibnamefont{Walker}}, eds.,
  \emph{\bibinfo{title}{International Linear Collider Reference Design Report,
  Volume 3: Accelerator}} (\bibinfo{publisher}{International Committe for
  Future Accelerators}, \bibinfo{year}{2007}),
  \bibinfo{note}{\texttt{http://www.linearcollider.org/}}.

\bibitem[{\citenamefont{Urban and Dugan}(2005)}]{Urban}
\bibinfo{author}{\bibfnamefont{J.}~\bibnamefont{Urban}} \bibnamefont{and}
  \bibinfo{author}{\bibfnamefont{G.}~\bibnamefont{Dugan}}, in
  \emph{\bibinfo{booktitle}{Proceedings PAC 2005}} (\bibinfo{publisher}{IEEE},
  \bibinfo{year}{2005}), p. \bibinfo{pages}{1880}.

\bibitem[{\citenamefont{{R. D. Ryne et al}}(2006)}]{MLI}
\bibinfo{author}{\bibnamefont{{R. D. Ryne et al}}}, in
  \emph{\bibinfo{booktitle}{Proceedings of the 2006 International Computational
  Accelerator Physics Conference, Chamonix, France}} (\bibinfo{year}{2006}),
  pp. \bibinfo{pages}{157--159}.

\bibitem[{\citenamefont{Dragt}(1999)}]{Dragt4}
\bibinfo{author}{\bibfnamefont{A.~J.} \bibnamefont{Dragt}}, in
  \emph{\bibinfo{booktitle}{Handbook of Accelerator Physics and Engineering}},
  edited by \bibinfo{editor}{\bibfnamefont{A.}~\bibnamefont{Chao}}
  \bibnamefont{and} \bibinfo{editor}{\bibfnamefont{M.}~\bibnamefont{Tigner}}
  (\bibinfo{publisher}{World Scientific Publishing}, \bibinfo{year}{1999}), pp.
  \bibinfo{pages}{78--84}.

\end{thebibliography}

\end{document}